\documentclass[english,a4paper,superscriptaddress]{revtex4}
\usepackage[dvips]{graphicx}
\usepackage{graphics}
\usepackage{epsfig}
\usepackage{amssymb}
\usepackage{amsmath}
\usepackage{babel}
\usepackage{float}

\begin{document}

\title{Contribution  of  a time-dependent metric on the dynamics of an interface between two immiscible electro-magnetically controllable Fluids}
\author{Q. Vanhaelen\footnote{corresponding author, email address: quentin.vanhaelen@ulb.ac.be}\footnote{Boursier F.R.I.A}}
\affiliation{Statistical and Plasma Physics, Universit\'e Libre de Bruxelles, Campus
Plaine, CP 231, B-1050 Brussels, Belgium}
\author{M. Hennenberg}
\affiliation{Microgravity Research Center, Fac. Sc. Appl.,Universit\'e Libre de Bruxelles, Av. F.D. Roosevelt,  B-1050 Brussels, Belgium}
\author{S. Slavtchev}
\affiliation{Insitute of Mechanics, Bulgarian Acad. of Sciences, Acad. G. Bontchev Str., Bl.4, 1113 Sofia, Bulgaria}
\author{B. Weyssow}
\affiliation{Statistical and Plasma Physics, Universit\'e Libre de Bruxelles, Campus
Plaine, CP 231, B-1050 Brussels, Belgium}

\begin{abstract}
\noindent
We consider the case of a deformable material interface between two immiscible moving media, both of them being magnetiable. The time dependence of the metric at the interface introduces a  non linear term, proportional to the mean curvature, in the surface dynamical equations of mass momentum and angular momentum.  We take into account the effects of that term also in the singular magnetic and electric fields inside the interface which lead to the existence of currents and charges densities through the interface, from the derivation of the Maxwell equations inside both bulks and the interface. Also, we give the expression for the entropy production and of the different  thermo-dynamical fluxes. Our results enlarge previous results from other theories where the specific role of the time dependent surface metric was insufficiently stressed.
\newline

\noindent \textbf{Keywords:} interface dynamics, ferrofluid, electro magnetic effects, time dependent curvature effects
\end{abstract}

\maketitle
\section{Introduction}
\noindent
A lot of work has been already done on the study of the dynamics of a interface between two continuous immiscible media (gaseous/liquid or liquid/liquid). From a macroscopic point of view, the thickness of the interface is generally assumed to be negligible, with respect to its extent. A certain number of methods deal with this problem. The study of  the interface between two media would be quite simple if one could consider the interface as a mathematical boundary problem. In that very first approach, the interface is not yet considered  as a "real" physical phase by itself since it doesn't have any intrinsic quantity such as mass density, current density, or momentum density. If the electo-magnetic field is taken into account, one have to add the usual boundary conditions for the normal and tangential components at the interface between both bulks like for a ferrofluid in the analysis of the Rosensweig instability \cite{rr}.  \vskip1.0em
\noindent Even if very useful, this first approach, hides the true physical nature of the interface. Already, for pure fluids, the Laplace law  expresses the jump of momentum balance along the interface to be equal to the surface tension times the mean curvature. Surface tension is the free energy of the surface: any real interface has a physical character leading to a much more complex problem. For example, how to take into account precisely a surface mass density, a surface flow ?And in the case of magnetizable media how to allow for the existence of a singular current between the two phases. To introduce explicitly a finite thickness is not a trivial matter. It leads to very tedious calculations \cite{ra,rb,rc,rd,re}. Indeed, to use this, for the electric fluid, see \cite{ib}, or for the magnetic fluid without electrical effect, see \cite{ra}$-$\cite{re} becomes very difficult, due to the expansion in series of the dimensionless thickness, as for example for a conducting fluid with fully electro-magnetic effects. We will not follow this development any further. Thus, we need another approach where the interfacial physical quantities are introduced and are singular in the following way: they exist \textit{inside} the interface and are meaningful as far as the interface exists. They have no existence \textit{normally} to the surface. Now the interface is a two-dimensional phase described in terms of intrinsic physical quantities for which dynamical balance equations have to be established, as is already the case for the adjacent bulk phases. This macroscopic  as well as mesoscopic approaches permit still to ignore explicitly, but up to a certain point  \cite{gapr}, the thickness of the interface \cite{bv},  even though it is different from zero, from a microscopic point of view. Indeed, we will suppose that any characteristic length is much larger than the characteristic width of the interface \cite{gapr,bb, ib}
\vskip1.0em\noindent
A first entirely macroscopic approach was developed systematically in \cite{ia,ic,id}, in terms of continuous fluid mechanics. The dynamical equations (mass conservation, energy conservation and momentum conservation) for the interface are obtained, using a 2 D physical model of the interface corresponding to the 3 D one describing the bulk phase. The method  consists in choosing a sample volume and in calculating for each physical entity its balance across the volumic sample, where the interface introduces  a discontinuity whenever the integration volume contains part of the interface. The integration introduces fluxes  of any physical quantities which have to be exchanged between each bulk at their common interface. This procedure leads to two sets of balance equations: one for the bulk phases which is of course the usual one describing a single volumic phase and another one describing the interface itself. But that last takes into account the adjacent bulks since they are responsible for the fluxes crossing the surface. Those surface balance equations introduce a new term coming from the time evolution of the geometry of the deformable interface. This term appears already in the Laplace law, through the mean curvature. But in other surface balance equations, a term proportional to the mean curvature times the flux of the surface physical quantity along the surface appears. It reflects the local change in the metric of the surface itself and it is a highly non linear term which does not intervene if the interface is a pure discontinuity without physical characteristics. This fully "integral" approach has been used to establish the dynamics of the mass surfactant on a interface layer, see \cite{if} and \cite{ig} and also for multi component system cf.\cite{ie}.
\vskip1.0em
\noindent
Another mathematical derivation has been proposed by D. Bedeaux, A.M. Albano and P.A. Wolff \cite{bb,bc}, and was extended to polarizable cases by Albano and coworkers  \cite{bd, be, bf, bg, bh} . Their aim was to build a suitable theoretical framework to apply non-equilibrium thermodynamics in fluid systems containing an interface and to relate it to statistical mechanics \cite{bg} . In \cite{bb} the mathematical basis has been done and  they apply the theory to a two fluids system with interface where singular energy density is taken into account; the extension to the multi-component system was done by Kovac, \cite{bc} who also considered the case of fluid with internal angular momentum, \cite{be}. However, this last author did not consider his model in conjunction with an imposed electromagnetic field. This author furthermore restricted his analysis to the flat undeformable interface\cite{bc}, since then the surface tension is independent from the choice of the dividing surface when one supposes local equilibrium \cite{prigo}. \vskip1.0em\noindent
The theoretical approach of Bedeaux and co. \cite{ba,bb,bc,be,bf,bg,bh} has been used for a system with electromagnetic effect: firstly without singular magnetic field in the interface, \cite{be} and with singular densities and current in \cite{bf,bg,bh}. But, each time, the balance of angular momentum has not been considered. The importance of the approach is how to treat the thickness: the interface position is considered as proportional to a delta function and the two bulk phases existence is taken into account formally by a Heaviside function, this formal decomposition has the advantage to suppress the problem related to the integration of the thickness. In  one of their first papers, Albano \cite{bf}discussed the very sensitive problem linked to the interface mean position \cite{gapr} and gave several reason to justify this method for a lot of systems. The change of shape of the surface is expressed by the mean curvature $\mathcal{H}$ which  appears explicitly only in the normal component of the jump of momentum at the interface. The change of metric expresses the local variation of distances along  the interface: expansion or contraction of the surface. It has to be taken into account, as well as local stretching \cite{if}. The non linear term related to the time dependent metric has been shown to exist in  previous works \cite{ia,ib,ic,id, ie,if}. In fact, it appears explicitly in the equations, but it is not sufficiently stressed in the original approach of Bedeaux and coworkers, leading perhaps to some misunderstandings. In conclusion, all three  approaches \cite {ia,ib,ic,id,ie,if,bren,gapr} or Gogosov and al.\cite{ra,rb,rc,rd,re}  and the one derived by Bedeaux et al. \cite{bd, be, bf, bg, bh} lead to the same surface balance equations
.\vskip1.0em\noindent
This paper is organized as follows: in section two, we recall the essential mathematical basis of the theory, since much of the detailed  derivations  are to be found in \cite{bb}.   In section three, the dynamical equations for the linear momentum and for the electro-magnetic field, in the interface and for the bulks phases are established. Indeed, our main aim is to extent the basic study of Rosensweig \cite{odenb} to the case where a free magnetisable material interface separates two immiscible  fluids, neglecting relativistic terms \cite{rosen,feld}. In section four we compute the governing equations for the total, internal and external angular momentum. In section five, we derive the constitutive equations for the magnetization and for the polarization. In section six, we establish the conservation equations inside the interface and the bulk for the kinetic energy, the rotational kinetic energy  and for the electro magnetic energy. In section seven, the corresponding equations for the internal total energy are obtained. However, in each case, we sill stress explicitly the role of the time dependent metric since  Bedeaux et al.\cite{bb} did not. In the next section, we discuss the entropy production and associated fluxes. We end up with the conclusion.
\section{Mathematical modeling of the interface}
\noindent
In this section, we present the mathematical basis to model the system, starting from the mathematical development  explicitly done in the past\cite{ba,bb,bc,bd}. We will review the general definitions of the different quantities to be used and  then, we  show how the time dependence of the curvature can be taken into account in this formalism, we will apply this new feature to the equation for the conservation of mass of the system.
\subsection{Mathematical Basis of the theory }
\noindent
Let us recall the presentation made in \cite{bb}, also used by \cite{bc}. For simplicity, we consider  two distinct and immiscible one-component fluids separated by a time dependent and spatially deformable  interface, $\mathcal{S}(t)$, defined by the function
$f(\vec{x},t) = 0$ such that $f(\vec{x},t)>0$ corresponds to the fluid I and that $f(\vec{x},t)<0$ defines the fluid II.  Each physical quantity, $\mathcal{A}$ is written as the sum of the two bulk phases quantities ($\mathcal{A}^{+}$ for the phase I, $\mathcal{A}^{-}$ for the phase II), and a singular quantity representing the interface itself, $\mathcal{A}^s$ as:
\begin{equation}\label{a}
\mathcal{A}(\vec{x},t) = \mathcal{A}^{-}\Theta^{-}+\mathcal{A}^{+}\Theta^{+}+\mathcal{A}^{s}\delta^{s}
\end{equation}
Equation (\ref{a}) introduces two new functions which are the most important mathematical tools to be used. To begin, the Heaviside step function, defined as follows:\[\Theta^{+}(f)= \begin{cases}
1 & \text{if } f>0\\
0 &\text{if } f\leq 0
\end{cases}
\qquad
\Theta^{-}(f)= \begin{cases}
0 & \text{if } f\geq0\\
1 &\text{if } f<0
\end{cases}
\]
The Heaviside step function enjoys a very important property :
\begin{equation}\label{b}
\frac{\partial \Theta^{\pm}}{\partial f} = \pm \delta(f)
\end{equation}
where $\delta(f)$ is the Dirac function. Since exists a direct relation between the Heaviside function and the delta function, the "surface delta function", $\delta^s$, which appears in (\ref{a}) will be expressed in terms of $\delta(f)$. To do that, Bedeaux et al. \cite{bb} consider the total area of the interface:
\begin{equation}\label{c}
\Xi(t) = \int\left|\vec{\nabla} f(\vec{x},t)\right|\delta(f)d\vec{x}
\end{equation}


\noindent
The normal vector at the interface, $\vec{n}$, pointing from bulk II into bulk I is defined as:
\begin{equation}\label{e}
\vec{n} = \left.\frac{\vec{\nabla} f}{\left|\vec{\nabla} f\right|}\right|_{f=0}
\end{equation}
From  expression (\ref{c}) and (\ref{e}), Bedeaux et al. \cite{bb} defined the function $\delta^s(f)$ as:
\begin{equation}\label{d}
\delta^s(\vec{x},t) = \left|\vec{\nabla} f(\vec{x},t)\right|\delta(f(\vec{x},t))
\end{equation}
That function $\delta^s(\vec{x},t)$ has an enviable property: one can restrict any integration of the volume that contains the interface of discontinuity , to an integration over the discontinuity surface $\mathcal{S}(t)$.
Using equations (\ref{b}, \ref{e}, \ref{d}), we have:
\begin{equation}\label{f}
\frac{\partial\Theta^{\pm}(f(\vec{x},t))}{\partial \vec{x}} = \pm \vec{n}\delta^s(\vec{x},t)
\end{equation}
\noindent
By definition, the function $f$ is equal to zero at the interface, so that we have:
\begin{equation}\label{g}
\frac{\partial f}{\partial t} = -\vec{u}^s.\nabla f = -u^s_n\left|\nabla f\right|, \qquad \mbox{for} \quad f = 0
\end{equation}
where $u^s$ is the velocity of an object \textit{intrinsic} to the surface and thus, $u^s_n = \vec{n}.\vec{u}^s$ is the interface velocity, taken normally to the interface. It can differ from $\vec{u}^{\pm}|_{f=0}.\vec{n}$ which is the normal component of the $\pm$ bulk velocity $\vec{u}^{\pm}$ along the surface, like for example in non equilibrium crystal growth
\cite{gg}. The difference $\displaystyle{[\vec{u}_{\pm|_{f(\vec{x},\,t)=0}}.\vec{n}-\vec{u}^s_n].\vec{n}}$ defines thus a relative velocity\cite{bc,be}\cite{ia}-,\cite{id}, at the interface and thus a flux towards or leaving it.
\vskip1.0em
\noindent
The time derivative of the Heaviside function is:
\begin{equation}\label{h}
\frac{\partial \Theta^{\pm}(f(\vec{x},t))}{\partial t} = \mp u^s_n\delta^s(\vec{x},t)
\end{equation}
Like Bedeaux \cite{bb}, one defines the total time derivative at the interface $f(\vec{x}, t)=0$ as follows:
\begin{equation}\label{i}
\frac{D_s}{Dt} = \frac{\partial}{\partial t}+\vec{u}^s.\nabla
\end{equation}
From last equations (\ref{h}, \ref{i}), Bedeaux et al. \cite{bb} derive the important results:
\begin{equation}\label{k}
\frac{D_s \theta^{\pm}(f)}{Dt} = \frac{D_s\delta^s}{Dt} = 0
\end{equation}
Another important property concerning the quantities on the interface is the following:
\begin{equation}\label{las}
\vec{n}.\vec{\nabla}\mathcal{A}^s = 0
\end{equation}
where $\mathcal{A}^s$ is a physical quantity intrinsic to the interface. Equation (\ref{las}) means that the quantities of the type $\mathcal{A}^s$ which are defined \textit{on the interface} cannot change in the normal direction to that interface: they can change only along the interface. In practice, thus, whatever $\mathcal{A}_s$ we consider, we have $\vec{\nabla}\mathcal{A}^s$ as $\vec{\nabla}_s\mathcal{A}^s$ with the definition of the operator $\vec{\nabla}_s = [(\textbf{I}-\vec{n}\vec{n}).\vec {\nabla}]$. \\

\noindent The properties (\ref{f},\ref{h},\ref{k},\ref{las}) will be very useful in any future calculation. Indeed, the delta and Heaviside functions lead to a simple definition of the interface. To compute the equations for the bulks and the interface are simpler than in another approach for which the definition of the different phases in terms of boundaries in the integrals leads to very tedious calculations even for situations simpler than those that we are going to deal with\cite{ra, rb, rc, rd, re}.
\subsection{Influence of the time dependent metric: Example of the continuity equation for the density}\noindent
The influence of the time dependent metric is actually well known \cite{ia}$-$\cite{ie}. For example, in \cite{if} and \cite{ig}, a simple derivation of the dynamics of the mass surfactant at the interface under the influence of the motion of the interface is given. It reflects the time dependence of the compressibility of the interface. However, the works \cite{ra}$-$\cite{ba} take explicitly into account only the effects of the time depending mean curvature. The change of shape is considered by Bedeaux et al., but it lies hidden in the formalism. Indeed, as we stressed in the introduction, Bedeaux and coworkers \cite{bb, bc, bd, be,bf}  take only into account the usual Laplace law, in the balance of momentum along the interface. In this section, we show how the time dependent surface metric can appear naturally, extending thus the mathematical formalism developed by Bedeaux and coworkers and finding back the results introduced by \cite{ia,ib,ic,id,ie,if,ig}, as part of our extension. \vskip1.0em\noindent
As an illustration of this approach, we consider the total mass conservation law and derive its formulation for each bulk and inside the interface $\mathcal{S}$. Let us consider the total density  $\rho_T$, using the decomposition (\ref{a}), we have:
\begin{equation}\label{ja}
\rho_T(\vec{x},t) =\rho^{+}(\vec{x},t)\Theta^{+} +\rho^{-}(\vec{x},t)\Theta^{-}+\rho^{s}(\vec{x},t)\delta^s
\end{equation}
where $\rho^{\pm}$ are the bulk densities of the $\pm$ bulk respectively extended up to the interface and $\rho^s$ might be the Gibbs superficial mass density, depending on the choice of the mean surface position. If we note the total volume of the system, $V_T$, we can rewrite it as $V_T = \mathcal{S}\oplus V^{+}\oplus V^{-}$. Supposing that the bulk phases are immiscible, we write the conservation of mass density in the integral form, in the absence of any chemical reaction:
\begin{equation}\label{jb}
\frac{D}{Dt}\int\rho_T dV_T = 0
\end{equation}
If we put the decomposition (\ref{ja}) in (\ref{jb}), we obtain:
\begin{equation}\label{jc}
\frac{D}{Dt}\int(\rho^s\delta^s)dV_T+\frac{D}{Dt}\int(\rho^{\pm}\theta^{\pm})dV_T = 0
\end{equation}
The first term in the LHS of(\ref{jc}) is proportional to $\delta^s$.The next term is proportional to the Heaviside function, which is non zero only for the corresponding bulk phases, then we can rewrite (\ref{jc}) as
\begin{equation}\label{jd}
\frac{D}{Dt}\int(\rho^s\delta^s)d\mathcal{S}+\frac{D}{Dt}\int(\rho^{+}\theta^{+})dV^{+}+\frac{D}{Dt}\int(\rho^{-}\theta^{-})dV^{-} = 0
\end{equation}
In the first term  of (\ref{jd}), the density $\rho^s$ is defined only on the interface, then we can write for this term
\begin{equation}
\frac{D_s}{Dt} = \frac{\partial}{\partial t}+\vec{u}_s.\vec{\nabla}
\end{equation}
Then the first term in eq.(\ref{jd}) is
\begin{equation}
\frac{D_s}{Dt}\int(\rho^s\delta^s)d\mathcal{S} = \int\frac{D_s}{Dt}(\rho^s\delta^s)dS+\int(\rho^s\delta^s)\frac{D_s}{Dt}dS
\end{equation}
and now using the second relation in eq.(\ref{k}) we have
\begin{equation}\label{fdg}
\frac{D_s}{Dt}\int(\rho^s\delta^s)d\mathcal{S} =\int(\frac{D_s\rho^s}{Dt})\delta^sdS+\int(\rho^s\delta^s)\frac{D_s}{Dt}dS
\end{equation}
\vskip1.0em\noindent  In the RHS of (\ref{fdg}), the last term is proportional to $\displaystyle{\frac{\partial d\vec{S}}{\partial t}}$ times $\delta^s$ . To evaluate it, we use the well known property \cite{ia,ib,ic,id,ie}
\begin{equation}\label{ghj}
\int\frac{\partial d\vec{S}}{\partial t} = \int\frac{1}{2 g}\frac{\partial g}{\partial t}d\vec{S}
\quad\mbox{so that}\quad\ \int\delta^s\mathcal{A}^s\,\frac{\partial d\vec{S}}{\partial t} = \int\delta^s\,\mathcal{A}^s\frac{1}{2 g}\frac{\partial g}{\partial t}d\vec{S}\end{equation}
\noindent
for any quantity $\mathcal{A}^s$ defined on the surface. By definition, $g = det(g_{ij})$, where $g_{ij}$ is the metric on the interface and where \cite{ia,ic,id,ie,aris}:
\begin{equation}\label{hjjk}
\frac{\partial g}{\partial t} = -4g\vec{u}^s.\vec{n}\mathcal{H}
\end{equation}
In last expression (\ref{hjjk}), we used the classical definition of the mean curvature $\mathcal{H} = -\frac{1}{2}\vec{\nabla}_s.\vec{n}$ (cf. \cite{bren, ia,ib,ic,id,ie,if,ig}). \vskip1.0em\noindent
Another way to obtain this important point rests on  the demonstration given by Stone \cite{if}, who  performed some mathematical transformations on the second term of the RHS of (\ref{fdg}) (see (\cite{if, bat}) for more details):
\begin{equation}\label{dens}
\frac{D_s}{Dt}d\vec{S} =[ d\vec{S}\vec{\nabla}].\vec{u}^s-(\vec{\nabla}\vec{u}^s).d\vec{S}
\end{equation}
where $d\vec{S} = \vec{n}.dS$. Taking the scalar product of eq.(\ref{dens}) with $\vec{n}$, one obtains
\begin{equation}
\frac{D_s}{Dt} dS = dS (\vec{\nabla}.\vec{u}^s)-(\vec{\nabla}\vec{u}^s.\vec{n}).\vec{n}dS
\end{equation}
and because
\begin{equation}
(\vec{\nabla}\vec{u}^s.\vec{n}).\vec{n} = (\vec{n}\vec{n}.\vec{\nabla}).\vec{u}^s
\end{equation}
we finally obtain \cite{bat}:
\begin{equation}
\frac{D_s}{Dt} dS  = dS\vec{\nabla}_s.\vec{u}^s
\end{equation}
Now, we can rewrite  (\ref{fdg}) as follows:
\begin{equation}
\frac{D_s}{Dt}\int(\rho^s\delta^s)d\mathcal{S} =\int\delta^s\left[\frac{\partial \rho_s}{\partial t}+\vec{u}_s.\vec{\nabla}_s\rho_s+\rho_s\vec{\nabla}_s.\vec{u}_s+\rho_s\vec{\nabla}_s(\vec{n}\vec{n}.\vec{u}^s) \right]dS
\label{fdgg}\end{equation}
we use the following result :
\begin{equation}
\rho_s\vec{\nabla}_s(\vec{n}\vec{n}.\vec{u}) = \rho_s(\vec{\nabla}_s.\vec{n})(\vec{n}.\vec{u})
\end{equation}
We finally obtain
\begin{equation}
\frac{D_s}{Dt}\int(\delta^s\rho^s)dS = \int\delta^s\left[\frac{\partial \rho_s}{\partial t}+\vec{u}_s.\vec{\nabla}_s\rho_s+\rho_s\vec{\nabla}_s.\vec{u}_s+\rho_s(\vec{\nabla}_s.\vec{n})(\vec{n}.\vec{u}^s)  \right]dS
\end{equation}
and taking into account the above definition of the mean curvature $\mathcal{H}$ , we have our main result:
\begin{equation}
\frac{D_s}{Dt}\int(\delta^s\rho^s)dS =\int\delta^s\left[\frac{\partial \rho_s}{\partial t}+\vec{u}_s.\vec{\nabla}_s\rho_s+\rho_s\vec{\nabla}_s.\vec{u}_s-2\mathcal{H}\rho_s(\vec{n}.\vec{u}^s) \right]dS
\label{fsur}\end{equation}

\noindent The last term in the RHS of (\ref{fsur}) introduces the mean curvature $\mathcal{H}$ times the normal component of the surface velocity $u^s_n$, enlarging the original demonstration given by Stone \cite{if}, since we are considering also the total mass flux coming or leaving the adjacent bulk phases. Furthermore, Stone did not give any clear definition of the velocity. This demonstration has been also used in the case of \cite{rd} and \cite{re}. But, these authors considered that $\vec{n}.\vec{u}^s = u^s_n$ the bulk velocity at the interface is the surface velocity. On the contrary, we will assume that the surface velocity \emph{at} the interface is $\vec{u}^s$, but this does not mean that \emph{at} the interface, the bulk velocity normal component $\vec{u}.\vec{n}$ calculated for $f(\vec{x}, t)=0$ has the value $\vec{n}.\vec{u}^s = u^s_n$ of the surface velocity. In our approach the "drift" terms of each phase which are proportional to $\vec{n}.[\vec{u}^{\pm}|_{f=0}-\vec{u}^s]$ are not neglected, as we will see just below.\\

\noindent Now, we assume that each phase is incompressible. Thus, the total volume of that phase remains constant: $\frac{DV^{+}}{Dt} =\frac{DV^{+}}{Dt}=0 $. The effects of the curvature are negligible inside the bulks and we have for the second term of (\ref{jc}), using the Reynolds transport theorem (see \cite{odenb}) and the relations (\ref{f}) and (\ref{h}) :
\begin{equation}
\frac{D}{Dt}\int(\rho^{\pm}\theta^{\pm})dV_T = \int dV_T\left[\frac{\partial\rho^{\pm}}{\partial t}\theta^{\pm}\mp\vec{n}.\vec{u}^s\rho^{\pm}\delta^s+\theta^{\pm}\vec{\nabla}.(\vec{u}^{\pm}\rho^{\pm})\pm\vec{n}.\vec{u}^{\pm}\rho^{\pm}\delta^{s}\right]
\label{ff}\end{equation}
We can rewrite last equation (\ref{ff}) as:
\begin{equation}
\frac{D}{Dt}\int(\rho^{\pm}\theta^{\pm})dV_T =\int dV\left[\frac{\partial\rho^{\pm}}{\partial t}\theta^{\pm}+\theta^{\pm}\vec{\nabla}.(\vec{u}^{\pm}\rho^{\pm})\right]+ \int dS\left[\mp\vec{n}.\vec{u}^s\rho^{\pm}\delta^s\pm\vec{n}.\vec{u}^{\pm}\rho^{\pm}\delta^{s}\right]
\label{fvol}\end{equation}
Thus summing (\ref{fsur}) and (\ref{fvol}), we obtain the explicit form of the integral equation (\ref{jd}). But, we know that the coefficients of $\delta^s$ and  $\Theta^{\pm}$ must vanish separately (cf.\cite{bb}), and thus we obtain the following results for the bulks phases:
\begin{equation}
\frac{\partial\rho^{\pm}}{\partial t}+\vec{\nabla}.(\vec{u}^{\pm}\rho^{\pm}) = 0
\label{jj}\end{equation}
and for the interface:
\begin{equation}\label{jf}
\frac{D_s\rho^s}{Dt}+\rho^s\vec{\nabla}_s.\vec{u}^s-2\rho^s u^s_n\mathcal{H}+ \rho^{+}\vec{n}.(\vec{u}^{+}-\vec{u}^s)-\rho^{-}\vec{n}.(\vec{u}^{-}-\vec{u}^s) = 0
\end{equation}
\noindent
This typifies thus the approach of Bedeaux and coworkers \cite{ba, bb,bc,bd,be}. Stone's physical explanation of the term proportional to $-\,2u^s_n\mathcal{H}$ was thus extended to consider the influence of the drift terms, $\rho^{+}\vec{n}.(\vec{u}^{+}-\vec{u}^s)-\rho^{-}\vec{n}.(\vec{u}^{-}-\vec{u}^s)$. In (\ref{jf}), $\vec{\nabla}_s.\vec{u}^s-2\,u^s_n\mathcal{H}$ expresses "a source like contribution resulting from local change in the area which are related to stretching and distortion"\cite{if}. These two terms appeared already in (\ref{fdgg}) since we are writing $\vec{u}^s$ as a the sum of two contribution: one is along the surface and the other is normal to the surface. The first one shows up as  $\vec{\nabla}_s.\vec{u}^s$ and expresses the compressibility of the surface along itself and the other $-2\,u^s_n\mathcal{H}$ is due to the change of shape of the interface. The equation for the conservation of the mass inside the interface (\ref{jf}) is exactly the expression derived in \cite{if},\cite{ig} or in the works by V.V. Gogosov and  co-authors \cite{ra,rb,rc,rd,re} if one supposes, as these last do, that the normal component of the bulk velocity at the interface is the velocity of the interface.
\section{Dynamical equations for the interface}
\noindent In this section we consider the derivation of the dynamical equations of our system, using the method described above to show explicitly the effect of the curvature. We suppose that both bulk phases are pure liquids since the multicomponent bulks with chemical reactions has been the subject of extensive studies\cite{bc,bd,ia,ib,ic,id,ie}. We will consider a system where  magnetic and electric effects exist. Thus, we define the electric field, $\vec{E}$, the magnetic field, $\vec{H}$, the magnetic induction, $\vec{B}$, with the relation: $\vec{B} = \mu_0(\vec{H}+\vec{M})$, where $\vec{M}$ is the magnetization of the system and $\mu_0$ the magnetic permeability; $\vec{D}$ is the displacement field which is related to the electric field through the relation $\vec{D} = \varepsilon_0\vec{E}+\vec{P}$, where $\vec{P}$ is the electric polarization and  $\varepsilon_0$ is the electric permitivity of the void. We define the charge density $\rho_e$ and the current density $\vec{j}$. However, for the time being, we will not consider terms linked to the semi relativistic approximation \cite{rosen,feld} and will neglect terms proportional to $\displaystyle{\frac{u^{\pm}}{c},\,\frac{u^s}{c}}$.  The Maxwell equations are written in MKSA units, like for Rosensweig \cite{odenb, rosen}
\subsection{Momentum balance equations}
\noindent Consider now the momentum balance equations starting with the Navier Stokes equations, written in an integral form, see \cite{odenb, rosen, feld}.  The general expression of the momentum balance is:
\begin{equation}\label{momt}
\frac{D}{Dt}\int(\rho\vec{u})dV_T+\frac{D}{Dt}\int(\vec{p}_{em})dV_T = \int[(\vec{\nabla}.\breve{\textbf{T}})+\rho\vec{F}_{ex}]dV_T
\end{equation}
where we have the total stress tensor $\breve{\textbf{T}}$ ( taking into account the pressure the viscous effects and an electromagnetic part \cite{odenb,rosen}), the vector for electromagnetic impulsion $\vec{p}_{em} =  \vec{D}\times\vec{B}$ and the external force of non electromagnetic origin (e.g.: gravity), $\vec{F}_{ex}$. Since we wish to extend the initial approach of  Rosensweig \cite{odenb, rosen} to the present case,  we are using like him the Minkowsky expression for the electromagnetic momentum, while Albano and co \cite{bd,bf,bg,bh} , or Felderhof and Kroh \cite{feld} took the Abraham form $\frac{1}{c^2}\,\vec{E}\times\vec{H}$ in their derivation.
\vskip1.0em
\noindent
Like we did for the density (\ref{ja}), we use the decomposition (\ref{a}) for the momentum, the tensor and the external force, replacing  in (\ref{a}) $\mathcal{A}(\vec{x},t)$  by $\rho\vec{u},\, \vec{p}_{em},\,\breve{\textbf{T}},\,\rho\vec{F}_{ex}$ respectively so that (\ref{momt}) becomes:
\begin{eqnarray}\begin{array}{c}
\displaystyle{\frac{D}{Dt}\int\left[\rho^{+}\vec{u}^{+}\Theta^{+}+\rho^{-}\vec{u}^{-}\Theta^{-}\right]dV_T+\frac{D}{Dt}\int(\rho^s\vec{u}^s\delta^s)dS+\frac{D}{Dt}\int\left[\vec{p}^{\,\,+}_{em}\Theta^{+}+\vec{p}^{\,\,-}_{em}\Theta^{-}\right]dV_T+\frac{D}{Dt}\int(\vec{p}^{\,\,s}_{em}\delta^s)dS =}
\\\\
\int\vec{\nabla}.(\breve{\textbf{T}}^{-}\Theta^{-}+\breve{\textbf{T}}^{+}\Theta^{+}+\breve{\textbf{T}}^s\delta^{s})dV_T+\int(\rho^{-} \vec{F}^{-}_{ex}\Theta^{-}+\rho^{+} \vec{F}^{+}_{ex}\Theta^{+}+\rho^{s} \vec{F}^{s}_{ex}\delta^s)dV_T\end{array}\label{moto}
\end{eqnarray}
\subsubsection{Bulks and surface momentum}
\noindent
To begin, let us calculate the three first terms appearing in the left hand side of last equation (\ref{moto}), that is we develop the time derivatives for the bulks and the surface momentum. We have
\begin{eqnarray}\begin{array}{c}\label{mob1}
\displaystyle{\frac{D}{Dt}\int\left[\rho^{+}\vec{u}^{+}\Theta^{+}+\rho^{-}\vec{u}^{-}\Theta^{-}\right]dV_T = \int\left[\frac{\partial \rho^{+}\vec{u}^{+}\Theta^{+}}{\partial t}+\vec{\nabla}.(\rho^{+}\vec{u}^{+}\vec{u}^{+}\Theta^{+})+\frac{\partial \rho^{-}\vec{u}^{-}\Theta^{-}}{\partial t}+\vec{\nabla}.(\rho^{-}\vec{u}^{-}\vec{u}^{-}\Theta^{-})\right]}
\\\\
\displaystyle{=\int\left[\frac{\partial \rho^{+}\vec{u}^{+}}{\partial t}+\vec{\nabla}.(\rho^{+}\vec{u}^{+}\vec{u}^{+})\right]\Theta^{+}dV^{+}+\int\left[\frac{\partial \rho^{-}\vec{u}^{-}}{\partial t}+\vec{\nabla}.(\rho^{-}\vec{u}^{-}\vec{u}^{-})\right]\Theta^{-}dV^{-}}
\\\\
\displaystyle{ +\int\left[\rho^{+}\vec{u}^{+}\vec{n}.(\vec{u}^{+}-\vec{u}^s)-\rho^{-}\vec{u}^{-}\vec{n}.(\vec{u}^{-}-\vec{u}^s)\right]\delta^sdS}\end{array}
\end{eqnarray}
 \noindent
where we used the properties of the Heaviside and delta function (\ref{h},\ref{i},\ref{k},\ref{las}). \vskip1.0em\noindent
The next term  to be calculated at the interface is:
\begin{eqnarray}\begin{array}{c}
\displaystyle{\frac{D}{Dt}\int[\rho^s\vec{u}^s]\delta^sdS = \int\left[\frac{D(\rho^s\vec{u}^s)}{Dt}\delta^s\right]dS+\int\left[\rho^s\vec{u}^s\delta^s\right]\frac{DdS}{Dt} }
\\\\
\displaystyle{ = \int\left[\frac{\partial(\rho^s\vec{u}^s)}{\partial t}+\vec{\nabla}_s.(\rho^s\vec{u}^s\vec{u}^s)-2\mathcal{H}(\vec{n}.\vec{u}^s)\rho^s\vec{u}^s\right]\delta^sdS }\end{array}
\label{mob2}\end{eqnarray}\noindent
This result follows directly from the method defined in (\ref{ghj}), which was used to obtain  (\ref{fsur}).
\subsubsection{Electromagnetic impulsion}\noindent
Let's consider now the term for the electromagnetic impulsion. Using (\ref{ghj}), we obtain the following expression:
\begin{eqnarray}\begin{array}{c}
\displaystyle{\frac{D}{Dt}\int \vec{p}^{\,\,\pm}_{em}\Theta^{\pm}dV_T+\frac{D}{Dt}\int \vec{p}^{\,\,s}_{em}\delta^sdS =}
\\\\
\displaystyle{ = \int\left[\frac{\partial (\vec{p}^{\,\,\pm}_{em}\Theta^{\pm})}{\partial t}+\vec{\nabla}.(\vec{u}^{\pm}\vec{p}^{\,\,\pm}_{em}\Theta^{\pm})\right]dV_T+\int\left[\frac{\partial (\delta^s\vec{p}^{\,\,s}_{em})}{\partial t}+\vec{\nabla}.(\vec{u}^s\vec{p}^{\,\,s}_{em}\delta^s)\right]dS}
\\\\\displaystyle{ - \int 2\mathcal{H}(\vec{u}^s.\vec{n})\vec{p}^{\,\,s}_{em}\delta^s dS}
\end{array}\label{mob3}\end{eqnarray}
\subsubsection{Stress tensors}\noindent
Now, we will calculate the RHS of eq.(\ref{moto}) Following (\ref{a}), we have to calculate the stress tensor:
\begin{equation}\label{ahny}
\int{\vec{\nabla}.\breve{\textbf{T}}}dV_T = \int\left[\vec{\nabla}.(\breve{\textbf{T}}^{-}\Theta^{-})+\vec{\nabla}.(\breve{\textbf{T}}^{+}\Theta^{+})+\vec{\nabla}.(\breve{\textbf{T}}^s\delta^s)\right]dV_T
\end{equation}
At this stage, it is useful to define a new stress tensor $\widehat{\textbf{T}}$
\begin{equation}
\widehat{\textbf{T}} = \breve{\textbf{T}} - \vec{u}\vec{p}_{em}
\end{equation}
It is by definition, the total stress tensor minus the flow of the electromagnetic impulsion \cite{odenb},\cite{rosen}. This enables us to obtain a much simpler expression for the momentum balance, since it eliminates term linked to  $\vec{\nabla}.\vec{u}\vec{p}_{em}$. Then, Eq.(\ref{ahny}) gives :
\begin{equation}
\int{(\vec{\nabla}.\widehat{\textbf{T}}^{+})\Theta^{+}}dV^{+}+\int{(\vec{\nabla}.\widehat{\textbf{T}}^{-})\Theta^{-}}dV^{-}+\int{\left[\vec{\nabla}.\widehat{\textbf{T}}^s+\vec{n}.(\widehat{\textbf{T}}^{+}-\widehat{\textbf{T}}^{-})\right]\delta^s}dS+\int(\tilde{\textbf{T}}^s.\vec{\nabla}\delta^s)dV_T
\label{mob4}\end{equation}
Using (\ref{h}) and (\ref{k}) one obtains the following relations for the electromagnetic impulsion term:
\begin{equation}
\vec{p}^s_{em}\frac{\partial \delta^s}{\partial t} = -\vec{p}^s_{em}(\vec{u}^s.\vec{\nabla}\delta^s)
\quad\mbox{and}\quad
\frac{\partial (\vec{p}^{\,\,\pm}_{em}\Theta^{\pm})}{\partial t} = \frac{\partial\vec{p}^{\,\,+}_{em}}{\partial t}\Theta^{+}+\frac{\partial\vec{p}^{\,\,-}_{em}}{\partial t}\Theta^{-}-\vec{u}^s.\vec{n}(\vec{p}^{\,\,+}_{em}-\vec{p}^{\,\,-}_{em})\delta^s
\label{mob5}\end{equation}
Inserting all the previous results (\ref{mob1}), (\ref{mob2}), (\ref{mob3}), (\ref{mob4}), (\ref{mob5}) in (\ref{moto}), and taking into account that the term proportional in $\Theta^{\pm}$ and $\delta^s$ must vanish separately,  the momentum balance for each bulk is given by:
\begin{equation}\label{ert}
\frac{\partial \rho^{\pm}\vec{u}^{\pm}}{\partial t}+\vec{\nabla}.(\rho^{\pm}\vec{u}^{\pm}\vec{u}^{\pm})+\frac{\partial\vec{p}^{\pm}_{em}}{\partial t} = \vec{\nabla}.\widehat{\textbf{T}}^{\pm}+\rho^{\pm}\vec{F}^{\pm}_{ex}
\end{equation}
while for the interface, its specific momentum balance reads
\begin{eqnarray}\begin{array}{c}
\displaystyle{\frac{\partial \rho^{s}\vec{u}^{s}}{\partial t}+\vec{\nabla}.(\rho^{s}\vec{u}^{s}\vec{u}^{s})+\frac{\partial\vec{p}^{s}_{em}}{\partial t} -2\mathcal{H}(\vec{u}^s.\vec{n})(\rho^s\vec{u}^s+\vec{p}^s_{em})+(\vec{u}^s.\vec{n})(\vec{p}^{\,-}_{em}-\vec{p}^{\,+}_{em})=}
\\\\
\displaystyle{\vec{n}.(\widehat{\textbf{T}}^{+}-\widehat{\textbf{T}}^{-})+\rho^{-}\vec{u}^{-}(\vec{u}^s-\vec{u}^s).\vec{n}-\rho^{+}\vec{u}^{+}\vec{n}.(\vec{u}^{+}-\vec{u}^s) +\vec{\nabla}.\widehat{\textbf{T}}^{s}+\rho^{s}\vec{F}^{s}_{ex}}\end{array}
\label{eru}\end{eqnarray}
We have also some terms proportional to $\vec{\nabla}\delta^s$, so that on the normal components:
\begin{equation}\label{con}
\vec{p}^s_{em}\vec{u}^s.\vec{n}+\widehat{\textbf{T}}^s.\vec{n} = 0
\end{equation}
\subsubsection{Explicit form of the stress tensor $\widehat{\textbf{T}}$}
\noindent
The tensor $\widehat{\textbf{T}}$ complete form, for each bulk has been introduced already by \cite{odenb,rosen,feld}:
\begin{equation}
\widehat{\textbf{T}} = \tilde{\textbf{T}}+\textbf{T}_{em}
\end{equation}
\noindent
Here, the tensor $\tilde{\textbf{T}}$ is, as follows:
\begin{equation}\label{abc}
\tilde{\textbf{T}} =\textbf{T}+ 2\xi\vec{\textbf{Z}}.(\vec{\omega}-\vec{\Omega})
\end{equation}
where $\xi$ is the vortex vorticity and  $\vec{\textbf{Z}}$ is the alternating polyadic corresponding to $\vec{1}_x\vec{1}_y\vec{1}_z\epsilon_{ijk}$ where
\begin{displaymath}
\epsilon_{ijk} =
\left\{\begin{array}{ll}
1& \quad(ijk = 123, 231, \quad\textrm{or}\quad312)\\
0&\quad (i=j, i=k, \quad\textrm{or}\quad j=k)\\
-1&\quad (ijk = 132, 213, \quad\textrm{or}\quad321)
\end{array}\right.
\end{displaymath}
In Eq. (\ref{abc}) we take into account of the effect of the vorticity (defined as $\vec{\omega} = \frac{1}{2}\vec{\nabla}\times\vec{u}$) as well as  the internal angular momentum contribution $\vec{\Omega}$. Those two last effects have not been considered in \cite{re}, or in \cite{bd,bg,bh,bf}. Kovac, \cite{be}, considered the intrinsic angular momentum of  individual particle rotation and thus spin, while ignoring possible electro-magnetic effects.  In (\ref{abc}), we will suppose that  \textbf{T} has the following form describing a Newtonian fluid:
\begin{equation}\label{hji}
\textbf{T} = -p\textbf{I}+2\eta\textbf{D}+\lambda(\vec{\nabla}.\vec{u})\textbf{I}\quad\mbox{where \textbf {I} is the unit tensor and}\quad \textbf{D} = \frac{1}{2}(\vec{\nabla}\vec{u}+(\vec{\nabla}\vec{u})^{T})
\end{equation}
\noindent
In the definition (\ref{hji}), $p$ is the hydrostatic pressure, $\eta$ is the shear viscosity, $\lambda$ the volume viscosity. We assume that both bulk phases are incompressible, so that $\vec{\nabla}.\vec{u}=0$
\vskip1.0em\noindent
We consider  also the usual form of the Maxwell tensor \cite{rr}:
\begin{equation}
\textbf{T}_{em} = -\frac{1}{2}\left[\varepsilon_0\vec{E}^2+\mu_0\vec{H}^2\right]\textbf{I}+\vec{D}\vec{E}+\vec{B}\vec{H}
\end{equation}
As before, all of the physical quantities (velocity field, $\vec{u}$, internal angular momentum,$\vec{\Omega}$,... ) are decomposed using the formula (\ref{a}). For the tensors, we take the following form:
\begin{equation}
\textbf{T}_{em} = \textbf{T}^{-}_{em}\theta^{-}+\textbf{T}^{+}_{em}\theta^{+}+\textbf{T}^s_{em}\delta^s
\end{equation}
with the definitions:
\begin{eqnarray}\begin{array}{c}
\textbf{T}^{\pm}_{em} = -\frac{1}{2}\left[\varepsilon_0(\vec{E}^2)^{\pm}+\mu_0(\vec{H}^2)^{\pm}\right]\textbf{I}+\vec{D}^{\pm}\vec{E}^{\pm}+\vec{B}^{\pm}\vec{H}^{\pm}
\\\\
\textbf{T}^s_{em} = -\frac{1}{2}\left[\varepsilon_0(\vec{E}^2)^{s}+\mu_0(\vec{H}^2)^{s}\right]\textbf{I}+\vec{D}^s\vec{E}^s+\vec{B}^s\vec{H}^s
\end{array}
\end{eqnarray}
We will define now a very simple form for the tensor  $\tilde{\textbf{T}}^s$. Even if the adjacent bulks are incompressible, the material interface is compressible in the case where there is no drift. Thus, the description of the interface as a kind of 2D Newtonian liquid is somewhat more complicated than the one for a bulk fluid. Basing ourselves on previous results, \cite{rr} the  derivation leads to the following expression for the stress tensor at the interface:
\begin{equation}
\tilde{\textbf{T}}^s = \sigma\textbf{I}_s+\tau^s+2\xi^s\vec{\textbf{Z}}.(\vec{\omega}^s-\vec{\Omega}^s)
\end{equation}
with the intrinsic surface stress $\tau^s = 2\eta^s\textbf{D}^s+\lambda^s(\vec{\nabla}_s.\vec{u}^s)\textbf{I}_s$ \cite{bren,ia,id, ie}. Let us note that the interface is a compressible 2D media if $u^s_n=\vec{u}.\vec{n}$. Then we have
\begin{equation}\label{agf}
\vec{\nabla}.\tilde{\textbf{T}}^s = 2\sigma\mathcal{H}\vec{n}+\vec{\nabla}_s\sigma+\vec{\nabla}_s\tau^s+\vec{\nabla}.(2\xi^s\vec{\textbf{Z}}.(\vec{\omega}^s-\vec{\Omega}^s) )
\end{equation}
The first term in the RHS of (\ref{agf}) is the  Laplace force,  proportional to the surface tension $\sigma$ times the mean curvature, $\mathcal{H}$. This term is present in all previous works (cf. \cite{bc},\cite{be}). The second term introduces the variation of surface tension along  the interface and expresses thus a Marangoni effect  while leaving free its precise physica-chemical origin \cite{rr,bc,be}. We are not aware that the surface angular momentum $(2\xi^s\vec{\textbf{Z}}.(\vec{\omega}^s-\vec{\Omega}^s))$, introduced in the last term on the RHS of (\ref{agf}), has ever been taken into account. Thus in a way,  the total stress tensor has a very simple but heavy mathematical form
\subsection{Maxwell's equations in a moving medium}
\noindent We use the same kind of approach to derive now the Maxwell equations for the interface and for the bulks phases. Thus applying to the Maxwell equations the decomposition (\ref{a}), we will obtain terms proportional to $\vec{\nabla}\delta^s$, in addition to the ones proportional to $\delta^s$ and $\Theta^{\pm}$ \cite{bf} and \cite{bg}. Like in previous derivations, this kind of terms will be related to the boundary conditions at the interface \cite{bb}. 
\subsubsection{No explicit time derivative in the Maxwell equations}
\noindent
For those two Maxwell equations, the derivation typified by (\ref{a}) is identical to the ones given in \cite{bf} and \cite{bg}, since the time dependency of the surface metric (\ref{ghj}) will then not intervene. We will show thus directly the results. \vskip1.0em
\noindent
A] Writing the equations in their local form the Gauss law:
\begin{equation}\label{bhnj}
\vec{\nabla}.\vec{D} = \rho_e
\end{equation}
Using the decomposition (\ref{a}) for the vector field $\vec{D}$ and the charge $\rho_e$, and using the properties for the time and spatial derivatives of the $\delta^s$ and $\Theta^{\pm}$ functions, we arrive  to the following results:
\begin{itemize}
\item terms proportional to $\theta^{\pm}$:
\begin{equation}
\varepsilon_0(\vec{\nabla}.\vec{E}^{\pm})+\vec{\nabla}.P^{\pm} = \rho^{\pm}_e
\end{equation}
\item  terms proportional to $\delta^s$:
\begin{equation}
\varepsilon_0\vec{n}.(\vec{E}^{+}-\vec{E}^{-})+\vec{n}.(\vec{P}^{+}-\vec{P}^{-})+\vec{\nabla}.(\varepsilon_0\vec{E}^s+\vec{P}^s) = \rho^s_e
\label {del1}\end{equation}
\item  terms proportional to $\vec{\nabla}\delta^s$:
\begin{equation}
\vec{n}.(\varepsilon_0\vec{E}^s+\vec{P}^s) = 0
\end{equation}
\end{itemize}
B] The next equation is the zero divergence constraint on the magnetic field:
\begin{equation}\label{bhn}
\vec{\nabla}.\vec{B} = 0
\end{equation}
which gives us the following equations for the interface and the bulks:
\begin{itemize}
\item   terms proportional to $\theta^{\pm}$:
\begin{equation}
\vec{\nabla}.\vec{B}^{\pm} = 0
\end{equation}
\item   terms proportional to $\delta^s$:
\begin{equation}
\vec{n}.(\vec{B}^{+}-\vec{B}^{-})+\vec{\nabla}.\vec{B}^s = 0
\label{del2}\end{equation}
\item   terms proportional to $\vec{\nabla}\delta^s$:
\begin{equation}
\vec{n}.\vec{B}^s = 0
\end{equation}
\end{itemize}
\subsubsection{The time derivative Maxwell equations}
\noindent Now, we will consider the time depending Maxwell equations,i.e. the Amp\'ere law and the Faraday law. Since we assume that both bulks and the surface are moving at maybe different velocities, we consider that they are seen from a  stationary fixed laboratory reference frame. We will calculate the local form of Maxwell's equation, starting from the total time derivative of the volume integrals or of the surface integrals and use thus in that process explicitly (\ref{fdgg}) \cite{stratton, panofsky}.\vskip1.0em
\noindent A] We begin with the Faraday law, whose integral form reads:
\begin{equation}\label{farad}
\int\vec{\nabla}\times\vec{E}.d\vec{S}_T = -\frac{D }{Dt}\int\vec{B}.d\vec{S}_T
\end{equation}
\noindent This expression is a relation between the curl of the electric field along  a arbitrary surface $S_T$ which encloses the volume sample $\pm$ and the interface between them, and the temporal derivative of the magnetic flux which goes through the surface $S_T$. As has become usual by now, we must take the total time derivative in the RHS of (\ref{farad}), since the surface $S(t)$deforms itself with time \cite{ stratton, panofsky}. We use the same kind of formal decomposition for the $dS_T$ as we did for the $dV_T$:
\begin{equation}\label{dec}
dS _T= dS^+\oplus dS^-\oplus dS^s
\end{equation}
\noindent As we did already for the $dV^{\pm}$, we neglect the temporal variation of the $dS^{\pm}$ surface elements, because both $dS^{\pm}$ are chosen as arbitrary surface elements, inside the corresponding $\pm$ bulk. The situation is quite different for the interface $dS^s$: for that last one, we must take into account its time variation.  Starting from eq.(\ref{farad}) we have using the decomposition (\ref{a}) and the relation (\ref{dec}):
\begin{equation}
\int\vec{\nabla}\times\left(\vec{E}^{+}\Theta^{+}+\vec{E}^{-}\Theta^{-}+\vec{E}^s\delta^s\right).d\vec{S} = -\frac{D }{D t}\int\left[\vec{B}^{+}\Theta^{+}+\vec{B}^{-}\Theta^{-}\right].d\vec{S}-\frac{D_s}{D t}\int\vec{B}^s\delta^s.d\vec{S}^s
\end{equation}
Applying (\ref{f}), (\ref{h}) in this equation leads to
\begin{equation*}
\int(\vec{\nabla}\times\vec{E}^{+})\Theta^{+}dS^{+}+\int(\vec{\nabla}\times\vec{E}^{-})\Theta^{-}dS^{-}+\int((\vec{E}^{-}-\vec{E}^{+})\times\vec{n}+\vec{\nabla}\times\vec{E}^s)\delta^sdS^{s}-\int(\vec{E}^s\times\vec{\nabla}\delta^s)dS = 
\end{equation*}
\begin{equation}\label{induc}
-\frac{D }{D t}\int\left[\vec{B}^{+}\Theta^{+}+\vec{B}^{-}\Theta^{-}\right].d\vec{S}-\int\left[\frac{D_s \vec{B}^{s}}{Dt}\right]\delta^{s}dS^{s}+\int(\vec{B}^s\delta^s).\frac{D d\vec{S}^s}{D t}
\end{equation}
Now, let us look at the first term in the right side of equation (\ref{induc}).  Following a procedure analog to the one used for the density conservation equation, we get:
\begin{equation}
\frac{D }{D t}\int\left[\vec{B}^{+}\Theta^{+}+\vec{B}^{-}\Theta^{-}\right].d\vec{S} = \int dS^{\pm}\left[\frac{D\vec{B}^{\pm}}{Dt}\Theta^{\pm}\right]+\int dS^s\left[\mp\vec{n}.\vec{u}^s\vec{B}^{\pm}\delta^s\pm\vec{n}.\vec{u}^{\pm}\vec{B}^{\pm}\delta^s\right]
\end{equation}
Now, taking into account that the terms proportional to $\delta^s,\Theta^{\pm}$ and $\vec{\nabla}\delta^s$ must vanish independantly from each other we will write the Faraday law for the bulk and for the interface as:
\begin{itemize}
\item  terms proportional to $\theta^{\pm}$:
\begin{equation}\label{ghu}
\frac{D \vec{B}^{\pm}}{D t} = - \vec{\nabla}\times\vec{E}^{\pm}
\end{equation}
Let us stress that the total time derivative appears quite naturally because the continuous bulk media are moving 
\item terms proportional to $\delta^s$:
\begin{equation}\label{dei}
\frac{D_s \vec{B}^s}{D t}\underbrace{-2\mathcal{H}\vec{B}^s\vec{u}^s.\vec{n}}_{\textrm{new term}}+\vec{B}^s\nabla_s.\vec{u}^s+\vec{B}^{+}\vec{n}.(\vec{u}^{+}-\vec{u}^s)-\vec{B}^{-}\vec{n}.(\vec{u}^{-}-\vec{u}^s) = - \vec{\nabla}\times\vec{E}^s-\vec{n}\times(\vec{E}^{+}-\vec{E}^{-})
\end{equation}
There appears now a drift flux of the inductions $B^{\pm}$, which did not show up in \cite{bf}, since they considered stationary bulks. 
\item terms proportional to $\vec{\nabla}\delta^s$:
\begin{equation}
\vec{n}\times\vec{E}^s =0
\end{equation}
\end{itemize}
The equation for the Amp\'ere law, written in the local form is:
\begin{equation}\label{laste}
\vec{\nabla}\times\vec{H} = \frac{\partial  \vec{D}}{\partial t}+\vec{j}
\end{equation}
However, we have again to go over to the integral form  and consider thus instead of (\ref{laste})\cite{stratton, panofsky}:
\begin{equation}
\int(\vec{\nabla}\times\vec{H})d\vec{S} = \frac{D}{D t}\int\vec{D}.d\vec{S}+\int\vec{j}.d\vec{S}
\end{equation}
Applying  the usual method to the Amp\'ere law, we obtain now the following equations:
\begin{itemize}
\item  terms proportional to $\theta^{\pm}$:
\begin{equation}\label{huk}
\vec{\nabla}\times\vec{H}^{\pm} = \frac{D \vec{D}^{\pm}}{D t}+\vec{j}^{\pm}
\end{equation}
\item   terms proportional to $\delta^s$:
\begin{equation*}
\frac{1}{\mu_0}\vec{n}\times(\vec{B}^{+}-\vec{B}^{-})+\frac{1}{\mu_0}\vec{\nabla}\times\vec{B}^s-\vec{n}\times(\vec{M}^{+}-\vec{M}^{-})-\vec{\nabla}\times\vec{M}^s = \varepsilon_0\frac{D_s \vec{E}^s}{D t}
\end{equation*}
\begin{equation}\label{bobu}
+\frac{D_s \vec{P}^s}{D t}+\vec{D}^s\nabla_s.\vec{u}^s+ \vec{D}^{+}\vec{n}.(\vec{u}^{+}-\vec{u}^s)-\vec{D}^{-}\vec{n}.(\vec{u}^{-}-\vec{u}^{s})+\vec{j}^s-\underbrace{2\mathcal{H}\vec{D}^s\vec{u}^s.\vec{n}}_{\textrm{new term}}
\end{equation}
\item  terms proportional to $\vec{\nabla}\delta^s$:
\begin{equation}
\vec{n}\times\vec{M}^s-\frac{1}{\mu_0}\vec{n}\times\vec{B}^s = 0
\end{equation}
\end{itemize}
\noindent The two first Maxwell's equations (\ref{del1}) and (\ref{del2}), describing the surface interaction with the  electro-magnetic field are exactly the same as in \cite{bg}. But for the two others (\ref{dei}) and (\ref{bobu}), we note the presence of a new term, proportional to the mean curvature and of a drift of induction and displacement. At this stage, we do not know how big these terms are. All we can say is that they are theoretically there. \vskip1.0em\noindent
Furthermore, Gogosov and coworkers of \cite{re} consider the situation of a ferrofluid without electric field, thus they consider only the second (\ref{del2}) and the last  Maxwell equations (\ref {huk}) deduced respectively from (\ref{bhn}) and in (\ref{laste}) where they neglect any electrical sources. They used a development in a power series of a small parameter, chosen to be the ratio of the width of the interface to a characteristic length measured along that interface which is supposed thus to be of a very large extent. Using such a perturbation method, they established the analog of the usual boundary condition for the induction and the magnetic field, but in the absence of a drift term, since they assumed continuity of the normal velocities along the surface. Gogosov's results \cite{re} show  a dependence on the mean curvature to appear in the boundary conditions, as a consequence of the deformation and the dynamics of the interface.\\
\subsubsection{Continuity equation for the charge density conservation}
\noindent In its integral form, the continuity equation for the charge density is:
\begin{equation}
\frac{D }{D t}\int(\rho_e)dV_T = -\int\vec{\nabla}.\vec{j}dV_T
\end{equation}
Using the decomposition (\ref{a}) for the current $\vec{j}$ and the charge $\rho_e$ with $V_T = \mathcal{S}\oplus V^{+}\oplus V^{-}$ as before, and using the properties (\ref{f}) and (\ref{h}), we obtain the following results:
\begin{itemize}
\item   terms proportional to $\theta^{\pm}$:
\begin{equation}
 \frac{D \rho^{\pm}_e}{Dt}+\vec{\nabla}.\vec{j}^{\pm}  =  0
\end{equation}
\item   terms proportional to $\delta^s$:
\begin{equation}
\frac{D_s \rho^s_e}{D t}+\rho^s_e\nabla_s.\vec{u}^s-2\mathcal{H}(\vec{u}.\vec{n})\rho_e+\vec{\nabla}.\vec{j}^s+\vec{n}.(j^{+}-\vec{j}^{-}) +\rho_e^+ \vec{n}.(\vec{u}^+-\vec{u}^s)- \rho_e^- \vec{n}.(\vec{u}^- -\vec{u}^s)= 0
\end{equation}
\item  terms proportional to $\vec{\nabla}\delta^s$:
\begin{equation}
\vec{n}.\vec{j}^s= 0
\end{equation}
\end{itemize}
\section{Equations for the angular momentum}
\subsection{Total angular momentum}
\noindent Following \cite{be,odenb,rosen,feld}, we define the total angular momentum of our system as :
\begin{equation}\label{kil}
\rho \vec{K} = \vec{r}\times(\rho\vec{u}+\vec{p}_{em})+\rho\,\vec{\Omega} = \rho\vec{L}+\rho\vec{\Omega}+\vec{r}\times\vec{p}_{em}
\end{equation}
We have already introduced the Poynting vector  $\vec{p}_{em} = \vec{D}\times\vec{B}$. The external angular momentum $\vec{L}$ due to fluid vorticity is defined as $\vec{L} = \vec{r}\times\vec{u}$, where $\vec{r}$ is the position vector and $\vec{\Omega}$ is the internal angular momentum (spin density) introduced by individual particle rotation\cite{be, odenb, mis}. From \cite{odenb} or \cite{be}(derived by Kovac in the absence of electro-magnetic fields), we know that the dynamical equation for the total angular momentum takes the following integral form:
\begin{equation}\label{tota}
\frac{D}{Dt}\int\rho\vec{K}dV = \int\vec{\nabla}.(\vec{r}\times\breve{\textbf{T}})dV+\int\vec{\nabla}.\textbf{Y}dV+\int\rho\mathcal{L}dV+\int\vec{r}\times\rho\vec{F}_{ex}dV
\end{equation}
\noindent where $\rho\mathcal{L}$ is the external torque applied by the system environment. The tensor $\textbf{Y}$ is called stress couple tensor by Rosensweig  \cite{rosen} and convective flux of angular momentum by Kovac \cite{be}. This tensor $\textbf{Y}$ and the tensor $\breve{\textbf{T}}$, using (\ref{abc}), are defined as follows (see for more details \cite{odenb,rosen} ):
\begin{eqnarray}\begin{array}{c}
\textbf{Y} = \eta\mathrm{'}(\vec{\nabla}\Omega +(\vec{\nabla}\Omega)^T)+2\xi\mathrm{'}\vec{\textbf{Z}}.(\vec{\nabla}\times\vec{\Omega})+\lambda\mathrm{'}(\vec{\nabla}.\vec{\Omega})\textbf{I}\\\\
\breve{\textbf{T}}=\tilde{\textbf{T}} +\textbf{T}_{em}+\vec{u}\vec{p}_{em}=\textbf{T}+ 2\xi\vec{\textbf{Z}}.(\vec{\omega}-\vec{\Omega})+\textbf{T}_{em}+\vec{u}\vec{p}_{em}
\end{array}\end{eqnarray}
\noindent Where $\eta\mathrm{'}$ is the shear coefficient of spin viscosity, $\lambda\mathrm{'}$ is the volume coefficient of spin viscosity and $\xi\mathrm{'}$ is the vortex spin vorticity. We will apply the usual decomposition (\ref{a}) for  the physical quantities $\rho\vec{K},\,\vec{p}_{em},\,\vec{\Omega},\,\breve{\textbf{T}},\,\textbf{Y}$, respectively.
We can rewrite (\ref{tota})  as follow (using $V_T = V^{+}\oplus V^{-}\oplus S$):
\begin{equation*}
\int\left(\frac{\partial\rho^{\pm}\vec{K}^{\pm}\Theta^{\pm}}{\partial t}+\vec{\nabla}.(\vec{K}^{\pm}\rho^{\pm}\vec{u}^{\pm}\Theta^{\pm})\right)dV+\int\left(\frac{D(\rho^{s}\vec{K}^s\delta^s)}{Dt}\right)dS+\int\left(\rho^s\vec{K}^s\delta^s\right)\frac{DdS}{Dt} =
\end{equation*}
\begin{equation*} \int\vec{\nabla}.(\vec{r}\times\breve{\textbf{T}}\Theta^{\pm})dV+\int\vec{\nabla}.(\vec{r}\times\breve{\textbf{T}}^s\delta^s)dV+\int\vec{\nabla}.(\textbf{Y}^{\pm}\Theta^{\pm})dV^{\pm}+\int\rho^{\pm}\Theta^{\pm}\mathcal{L}dV^{\pm}+\int\vec{r}\times\rho^{\pm}\Theta^{\pm}\vec{F}_{ex}dV^{\pm}+
\end{equation*}
\begin{equation}
\int\vec{\nabla}.(\textbf{Y}^s\delta^s)dV+\int\rho^s\delta^s\mathcal{L}dS+\int\vec{r}\times\rho^s\delta^s\vec{F}_{ex}dS
\end{equation}
This equation becomes:
\begin{equation*}
\int\left(\frac{\partial\rho^{\pm}\vec{K}^{\pm}\Theta^{\pm}}{\partial t}+\vec{\nabla}.(\vec{K}^{\pm}\rho^{\pm}\vec{u}^{\pm}\Theta^{\pm})\right)dV+\int\left(\frac{\partial \rho^{s}\vec{K}^s}{\partial t}+\vec{\nabla}.(\vec{u}^s\rho^s\vec{K}^s)-2\rho^s\mathcal{H}\vec{K}^s(\vec{u}^s.\vec{n})\right)\delta^sdS =
\end{equation*}
\begin{equation*} \int\vec{\nabla}.(\vec{r}\times\breve{\textbf{T}}\Theta^{\pm})dV+\int\vec{\nabla}.(\vec{r}\times\breve{\textbf{T}}^s\delta^s)dV+\int\vec{\nabla}.(\textbf{Y}^{\pm}\Theta^{\pm})dV^{\pm}+\int\rho^{\pm}\Theta^{\pm}\mathcal{L}dV^{\pm}+\int\vec{r}\times\rho^{\pm}\Theta^{\pm}\vec{F}_{ex}dV^{\pm}+
\end{equation*}
\begin{equation}
\int\vec{\nabla}.(\textbf{Y}^s\delta^s)dV+\int\rho^s\delta^s\mathcal{L}dS+\int\vec{r}\times\rho^s\delta^s\vec{F}_{ex}dS
\end{equation}
For the right side, we use the properties (\ref{f}) and (\ref{h}) to obtain the following results:
\begin{itemize}
\item   bulk phases:
\begin{equation}
\frac{\partial \rho^{\pm}\vec{K}^{\pm}}{\partial t}+\vec{\nabla}.(\rho^{\pm}\vec{K}^{\pm}\vec{u}^{\pm}) = \vec{\nabla}.(\vec{r}\times\breve{\textbf{T}}^{\pm}+\textbf{Y}^{\pm})+\rho^{\pm}\vec{r}\times\vec{F}_{ex}+\rho^{\pm}\mathcal{L}
\label{kova}\end{equation}
This is exactly the equation for the angular momentum used by Rosensweig \cite{odenb,rosen}
\item interface:
\begin{equation*}
\frac{\partial \rho^{s}\vec{K}^{s}}{\partial t}+\vec{\nabla}.(\rho^{s}\vec{K}^{s}\vec{u}^{s})-2\mathcal{H}(\vec{u}^s.\vec{n})\rho^{s}\vec{K}^{s} = \vec{n}.\vec{u}^s(\rho^{+}\vec{K}^{+}-\rho^{-}\vec{K}^{-})-\vec{n}.(\rho^{+}\vec{K}^{+}\vec{u}^{+}-\rho^{-}\vec{K}^{-}\vec{u}^{-})+\vec{\nabla}.(\vec{r}\times\breve{\textbf{T}}^s+\textbf{Y}^s)
\end{equation*}
\begin{equation}
+\vec{n}.(\vec{r}\times(\breve{\textbf{T}}^{+}-\breve{\textbf{T}}^{-}))+\vec{n}.(\textbf{Y}^{+}-\textbf{Y}^{-})+\rho^{s}\mathcal{L}+\rho^s\vec{r}\times\vec{F}_{ex}
\label{kovab}\end{equation}
\item  boundary conditions (terms proportional to$\vec{\nabla}\delta^s$):
\begin{equation}
\vec{n}.(\vec{r}\times(\tilde{\textbf{T}}^s+\textbf{T}^s_{em}+\vec{u}^s\vec{p}^s_{em})+\textbf{Y}^s) = 0
\end{equation}
\end{itemize}
The two last equations extend the one found by Kovac \cite{be} to a deformable interface (which might be flat) and in conjunction with an electro-magnetic field 
\subsection{External angular momentum}
\noindent The equations for the external angular momentum $\vec{L} = \vec{r}\times\vec{u}$ for each phase are obtained directly from the corresponding  equations  (\ref{ert}) and(\ref{eru}) as a starting point. Taking the vectorial product of (\ref{ert}) and(\ref{eru}) by $\vec{r}$, the resulting equation is:
\begin{equation}
\frac{\partial\rho^{\pm}\vec{L}}{\partial t}+\vec{\nabla}.(\rho^{\pm}\vec{L}^{\pm}\vec{u}^{\pm}) =\vec{r}\times\vec{\nabla}.(\tilde{\textbf{T}}^{\pm}+\textbf{T}^{\pm}_{em})-\vec{r}\times\frac{\partial\vec{p}^{\pm}_{em}}{\partial t}+\vec{r}\times(\rho^{\pm}\vec{F}_{ex})
\end{equation}
But, using the property \cite{rr}:
\begin{equation}\label{appro}
\vec{r}\times(\vec{\nabla}.\textbf{T}) = -\vec{\nabla}.(\textbf{T}\times\vec{r})+\vec{\textbf{Z}}:\textbf{T}
\end{equation}
we obtain the following:
\begin{equation}
\frac{\partial\rho^{\pm}\vec{L}}{\partial t}+\vec{\nabla}.(\rho^{\pm}\vec{L}^{\pm}\vec{u}^{\pm}) = - \vec{\nabla}.(\tilde{\textbf{T}}^{\pm}\times\vec{r}+\textbf{T}^{\pm}_{em}\times\vec{r})+\vec{\textbf{Z}}:(\tilde{\textbf{T}}^{\pm}+\textbf{T}^{\pm}_{em})-\vec{r}\times\frac{\partial\vec{p}^{\pm}_{em}}{\partial t}+\vec{r}\times(\rho^{\pm}\vec{F}_{ex})
\label{kovac}\end{equation}
For the interface we start from the equation:
\begin{equation*}
\frac{\partial \rho^s\vec{L}^s}{\partial t}+\vec{\nabla}.(\rho^s\vec{L}^s\vec{u}^s)-\vec{r}\times\left(2\rho^s\mathcal{H}(\vec{u}^s.\vec{n})\vec{u}^s\right) = \vec{r}\times(\vec{\nabla}.\tilde{\textbf{T}}^s)+\rho^{-}\vec{L}^{-}\vec{n}.(\vec{u}^{-}-\vec{u}^s)-\rho^{+}\vec{L}^{+}\vec{n}.(\vec{u}^{+}-\vec{u}^{s})+\vec{r}\times(\vec{n}.(\tilde{\textbf{T}}^{+}-\tilde{\textbf{T}}^{-}))+
\end{equation*}
\begin{equation}
\vec{r}\times\left(\vec{n}.(\textbf{T}^{+}_{em}-\textbf{T}^{-}_{em})\right)+\vec{r}\times(\vec{\nabla}.\textbf{T}^s_{em})+\vec{n}.\vec{u}^s(\vec{r}\times\vec{p}^{+}_{em}-\vec{r}\times\vec{p}^{-}_{em})-\vec{r}\times\frac{\partial\vec{p}^s_{em}}{\partial t}+\vec{r}\times(\rho^s\vec{F}_{ex})
\end{equation}
Using (\ref{appro}) and the relation
\begin{equation}
\vec{r}\times(\vec{n}.\textbf{T}) = -(\vec{n}.\textbf{T})\times\vec{r}
\end{equation}
one obtains finally:
\begin{equation*}
\frac{\partial \rho^s\vec{L}^s}{\partial t}+\vec{\nabla}.(\rho^s\vec{L}^s\vec{u}^s)-2\mathcal{H}(\vec{u}^s.\vec{n})\rho^s\vec{L}^s =
\end{equation*}
\begin{equation*}
\rho^{-}\vec{L}^{-}\vec{n}.(\vec{u}^{-}-\vec{u}^s)-\rho^{+}\vec{L}^{+}\vec{n}.(\vec{u}^{+}-\vec{u}^{s}) - \vec{\nabla}.(\tilde{\textbf{T}}^s\times\vec{r})+\vec{\textbf{Z}}:\tilde{\textbf{T}}^s-\left(\vec{n}.(\tilde{\textbf{T}}^{+}-\tilde{\textbf{T}}^{-})\right)\times\vec{r}-
\end{equation*}
\begin{equation}
\vec{\nabla}.(\textbf{T}^s_{em}\times\vec{r})+\vec{\textbf{Z}}:\textbf{T}^s_{em}-\left(\vec{n}.(\textbf{T}^{+}_{em}-\textbf{T}^{-}_{em})\right)\times\vec{r}-\vec{n}.\vec{u}^s(\vec{p}^{+}_{em}-\vec{p}^{-}_{em})\times\vec{r}-\vec{r}\times\frac{\partial\vec{p}^s_{em}}{\partial t}+\vec{r}\times(\rho^s\vec{F}_{ex})
\label{kovad}\end{equation}
together with the following normal boundary condition:
\begin{equation}
\vec{r}\times\left(\vec{n}.(\tilde{\textbf{T}}^s+\textbf{T}^s_{em})\right)+\vec{u}^s.\vec{n}(\vec{r}\times\vec{p}^s_{em} ) = 0
\end{equation}
\subsection{Internal angular momentum}
\noindent  Applying the B.A.M theoretical derivation, Kovac (see \cite{be}) considers a system including angular momentum effects, with external as well as internal angular momentum. But he does not take into account any electro-magnetic effects. Thus in his work \cite{be}, the total angular momentum reduces to: $\vec{K} = \vec{\Omega}+\vec{L}$, which simplifies the obtention of the equations for the internal angular momentum: he  substracts the equation (\ref{kova}) and (\ref{kovab}) for the external angular momentum, from the corresponding equation (\ref{kovac}) and (\ref{kovad}) for $\vec{K}$. \vskip1.0em\noindent
When electro-magnetic fields are present, the definition of the total angular momentum is different as we recalled just above (see (\ref{kil})\cite{odenb,rosen,feld}. Then, we follow an approach analogous to the one of R.Rosensweig in \cite{odenb} combining it with the BAM derivation to obtain an equation describing the internal angular momentum balance in the bulk phase and in the surface.  From the definition (\ref{kil}) for the total angular momentum, (\ref{tota}) becomes:
\begin{equation}\label{del}
\frac{D}{Dt}\int\left[\vec{r}\times\rho\vec{u}+\vec{r}\times\vec{p}_{em}+\rho\vec{\Omega}\right]dV = \int\vec{\nabla}.(\vec{r}\times\breve{\textbf{T}})dV+\int\vec{\nabla}.\textbf{Y}dV+\int\rho\mathcal{L}dV+\int\vec{r}\times\rho\vec{F}_{ex}dV
\end{equation}
We consider individually the different terms appearing in the left side of (\ref{del}). The first term is:
\begin{equation}
\frac{D}{Dt}\int\vec{r}\times\rho\vec{u}dV = \int\left[\rho^{\pm}\frac{D}{Dt}(\vec{r}\times(\vec{u}^{\pm}\Theta^{\pm}))\right]dV+\int\left[\frac{D}{Dt}(\vec{r}\times\rho^s\vec{u}^s\delta^s)\right]dS+\int\left[\vec{r}\times\rho^s\vec{u}^s\delta^s\right]\frac{DdS}{Dt}
\end{equation}
where we have by definition $\vec{u} = \frac{D\vec{r}}{Dt}$, so that $\vec{u}\times\frac{D\vec{r}}{Dt} = 0$. We obtain the following expression:
\begin{equation}
\frac{D}{Dt}\int\vec{r}\times\rho\vec{u}dV =\int\left[\vec{r}\times\rho^{\pm}\frac{D\vec{u}^{\pm}\Theta^{\pm}}{Dt}\right]dV+\int\vec{r}\times\left[\frac{\partial \rho^s\vec{u}^s\delta^s}{\partial t}+\vec{\nabla}.(\rho^s\vec{u}^s\delta^s)-2\mathcal{H}(\vec{u}^s.\vec{n})\rho^s\vec{u}^s\delta^s\right]dS
\end{equation}
Now, applying the usual BAM derivation (see (\ref{a})up to (\ref{ff}) and (\ref{fvol})),  the second term in the LHS of (\ref{del}) gives:
\begin{equation}
\frac{D}{Dt}\int\vec{r}\times\vec{p}_{em}dV = \frac{D}{Dt}\int\vec{r}\times(\vec{p}^{\pm}_{em}\Theta^{\pm})dV+\frac{D}{Dt}\int\vec{r}\times(\vec{p}^{s}_{em}\delta^{s})dS
\end{equation}
Using systematically the decomposition (\ref{appro}), last expression can be rewritten as follows:
\begin{equation*}
\frac{D}{Dt}\int\vec{r}\times\vec{p}_{em}dV = \int\left[-\vec{\textbf{Z}}:(\vec{u}^{\pm}\vec{p}^{\pm}_{em}\Theta^{\pm})+\vec{r}\times\left(\frac{\partial \vec{p}^{\pm}_{em}\Theta^{\pm}}{\partial t}+\vec{\nabla}.(\vec{u}^{\pm}\vec{p}_{em}^{\pm}\Theta^{\pm})\right)\right]dV
\end{equation*}
\begin{equation}
+\int\left[-\vec{\textbf{Z}}:(\vec{u}^s\vec{p}^{s}_{em}\delta^s)+\vec{r}\times\left(\frac{\partial\vec{p}^s_{em}\delta^s}{\partial t}+\vec{\nabla}.(\vec{u}^s\vec{p}^s_{em}\delta^s)\right)\right]dV-2\int\left(\mathcal{H}u^s_n\delta^s(\vec{r}\times\vec{p}^s_{em})\right)dS
\end{equation}
The decomposition (\ref{appro}) applies also to the right side of (\ref{del}), beginning with the first term, rewritten as:
\begin{equation}
\int\vec{\nabla}.(\vec{r}\times\breve{\textbf{T}})dV = \int\left[\vec{r}\times(\vec{\nabla}.\breve{\textbf{T}})-\vec{\textbf{Z}}:\breve{\textbf{T}}\right]dV
\end{equation}
\vskip1.0em\noindent 
We have
\begin{equation}
\int\vec{\nabla}.(\vec{r}\times\breve{\textbf{T}})dV = \int\left[\vec{r}\times\vec{\nabla}.(\breve{\textbf{T}}^{\pm}\Theta^{\pm})-\vec{\textbf{Z}}:(\breve{\textbf{T}}^{\pm}\Theta^{\pm})+\vec{r}\times\vec{\nabla}.(\breve{\textbf{T}}^s\delta^s)-\vec{\textbf{Z}}:(\breve{\textbf{T}}^s\delta^s)\right]
\end{equation}\noindent
If we put all the results in equation (\ref{del}), we can rewrite it as follows:
\begin{equation*}
\int\left[\vec{r}\times\left(\rho^{\pm}\frac{D}{Dt}(\vec{u}^{\pm}\Theta^{\pm})-\vec{\nabla}.(\widehat{\textbf{T}}^{\pm}\Theta^{\pm})+\frac{\partial(\vec{p}^{\pm}_{em}\Theta^{\pm})}{\partial t}-\rho^{\pm}\vec{F}_{ex}\Theta^{\pm}\right)\right]dV+
\end{equation*}
\begin{equation*}
\int\left[\vec{r}\times\left(\frac{\partial\rho^s\vec{u}^s\delta^s}{\partial t}+\vec{\nabla}.(\rho^s\vec{u}^s\delta^s)-2\mathcal{H}(\vec{u}^s.\vec{n})\rho^s\vec{u}^s\delta^s-\vec{\nabla}.(\widehat{\textbf{T}}^s\delta^s)-2\mathcal{H}u^s_n(\vec{p}^s_{em})\delta^s+\frac{\partial (\vec{p}^s_{em}\delta^s)}{\partial t}-\rho^s\vec{F}^s_{ex}\delta^s\right)\right]dV =
\end{equation*}
\begin{equation*}
-\frac{D}{Dt}\int(\rho^{\pm}\vec{\Omega}^{\pm}\Theta^{\pm})dV-\frac{D}{Dt}\int(\rho^{s}\vec{\Omega}^{s}\delta^{s})dS + \int\left[-\vec{\textbf{Z}}:(\widehat{\textbf{T}}^{\pm}\Theta^{\pm})+\vec{\nabla}.(\textbf{Y}^{\pm}\Theta^{\pm})+\vec{\nabla}.(\textbf{Y}^s\delta^s)+\rho^{\pm}\mathcal{L}\Theta^{\pm}\right]dV+
\end{equation*}
\begin{equation}
\int\left[\rho^s\mathcal{L}-\vec{\textbf{Z}}:\widehat{\textbf{T}}^s\delta^s\right]dS
\end{equation}
The RHS of last equation is equal to zero since it is the vectorial product of $\vec{r}$ by the momentum balance (\ref{eru}) and (\ref{ert}),for the bulk and for the interface. We get the following:
\begin{equation*}
\frac{D}{Dt}\int(\rho^{\pm}\vec{\Omega}^{\pm}\Theta^{\pm})dV+\frac{D}{Dt}\int(\rho^{s}\vec{\Omega}^{s}\delta^{s})dS =   \int\left[-\vec{\textbf{Z}}:(\widehat{\textbf{T}}^{\pm}\Theta^{\pm})+\vec{\nabla}.(\textbf{Y}^{\pm}\Theta^{\pm})+\vec{\nabla}.(\textbf{Y}^s\delta^s)+\rho^{\pm}\mathcal{L}\Theta^{\pm}\right]dV+
\end{equation*}
\begin{equation}
\int\left[\rho^s\mathcal{L}-\vec{\textbf{Z}}:\widehat{\textbf{T}}^s\delta^s\right]dS
\end{equation}
This is the equation for the internal momentum, which we were looking for. Then, the usual properties and procedure leads us to describe each bulk by:
\begin{equation}
\frac{\partial\rho^{\pm}\vec{\Omega}^{\pm}}{\partial t}+\vec{\nabla}.(\rho^{\pm}\vec{\Omega}^{\pm}\vec{u}^{\pm}) = \vec{\nabla}.\textbf{Y}^{\pm}-\vec{\textbf{Z}}:(\tilde{\textbf{T}}^{\pm}+\textbf{T}^{\pm}_{em})+\rho^{\pm}\mathcal{L}
\end{equation}
The equation at the interface has the form:
\begin{equation*}
\frac{\partial \rho^s\vec{\Omega}^s}{\partial t}+\vec{\nabla}.(\rho^s\vec{\Omega}^s\vec{u}^s)-2\mathcal{H}(\vec{u}^s.\vec{n})\rho^s\vec{\Omega}^s \end{equation*}\begin{equation}
= \vec{\nabla}.\textbf{Y}^s+\vec{n}.(\textbf{Y}^{+}-\textbf{Y}^{-})-\vec{\textbf{Z}}:\tilde{\textbf{T}}^s-\vec{\textbf{Z}}:\textbf{T}^s_{em} - \rho^{+}\vec{\Omega}^{+}\vec{n}.(\vec{u}^{+}-\vec{u}^s)+\rho^{-}\vec{\Omega}^{-}\vec{n}.(\vec{u}^{-}-\vec{u}^s)+\rho^s\mathcal{L}
\end{equation}
and the boundary conditions are
\begin{equation}
\vec{n}.\textbf{Y}^s =0
\end{equation}
\section{Constitutive equations for the relaxation processes}
\noindent To close the system of equations, we need to have dynamical equations for the magnetization, $\vec{M}$ and for the electric polarization $\vec{P}$. For example, in the case of a ferrofluid, one can consider it as a homogeneous one component stable solution \cite{rr,rosen}. However for such a system oscillating far from its stationary state, the collinearity assumption between the field $\vec{H}$ and the magnetization $\vec{M}$ does not hold and the equation of state of the form $\vec{M} = \vec{M}(T, \vec{H})=\chi\,\vec{H}$  (where $\chi$ is a constant), is not adequate anymore \cite{rosen,odenb,feld,mis}. It becomes mandatory to have  a dynamical equation for both quantities $\vec{M}$ and $\vec{P}$ in order to describe the relaxation process which occurs in this kind of situation. Although a lot of works has been developed to obtain these equations, mostly in relationship to the molecular origin of the relaxation process \cite{odenb,mis}, there is no real agreement on what is exactly the right equation to use at the macroscopic level. This is why, we will consider the most common phenomenological  equations for both polarization and magnetization relaxation processes,  appearing in the literature.
\subsection{Relaxation equation for the magnetization}
\noindent For the modeling of the relaxation process of the magnetization, we consider the Shliomis relaxation relationship, \cite{mis}, generalized for compressible fluids \cite{rosen}:
\begin{equation}
\frac{\partial\vec{M}}{\partial t}+\vec{\nabla}.(\vec{u}\vec{M}) = \vec{\Omega}\times\vec{M}+\frac{1}{\tau_m}(\vec{M}_{0}-\vec{M})
\end{equation}
where $\tau_m$ is the magnetic relaxation time constant and  $\vec{M}_{0}$ is the value of the magnetization at the equilibrium given by the usual  Langevin equation \cite{rr}. We assume that the value of these two quantities can be different for each phase of the system. Using the usual BAM derivation, we obtain then:\\
\begin{itemize}
\item For the bulk:
\begin{equation}\label{mre}
\frac{\partial\vec{M}^{\pm}}{\partial t}+\vec{\nabla}.(\vec{u}^{\pm}\vec{M}^{\pm}) = \vec{\Omega}^{\pm}\times\vec{M}^{\pm}+\frac{1}{\tau^{\pm}_m}(\vec{M}^{\pm}_{0}-\vec{M}^{\pm})
\end{equation}
\noindent We have considered the generalized form of the equation just for mathematical easiness. Nevertheless, in this work, we consider the bulk phase as a incompressible medium, then the equation (\ref{mre}) can be simplified as follow:
\begin{equation}
\frac{D\vec{M}^{\pm}}{D t}= \vec{\Omega}^{\pm}\times\vec{M}^{\pm}+\frac{1}{\tau^{\pm}_m}(\vec{M}^{\pm}_{0}-\vec{M}^{\pm})
\end{equation}
\item For the interface:
\begin{equation}
\frac{\partial\vec{M}^s}{\partial t}+\vec{\nabla}.(\vec{u}^s\vec{M}^s)-2\mathcal{H}(\vec{u}^s.\vec{n})\vec{M}^s + \vec{M}^{+}\vec{n}.(\vec{u}^{+}-\vec{u}^s)-\vec{M}^{-}\vec{n}.(\vec{u}^{-}-\vec{u}^s)= \vec{\Omega}^s\times\vec{M}^s+\frac{1}{\tau^s_m}(\vec{M}^s_{0}-\vec{M}^s)
\end{equation}
\end{itemize}
\noindent We  notice that we do not have an equation proportional to $\nabla\delta^s$. It means that the boundary conditions for the magnetization are completely determined by the boundary conditions obtained from the Maxwell equations.
\subsection{Relaxation equation for the polarization}
\noindent For the polarization we can use a similar equation as the one for the magnetic relaxation process (see also \cite{rosen}):
\begin{equation}
\frac{\partial\vec{P}}{\partial t}+\vec{\nabla}.(\vec{u}\vec{P}) = \frac{1}{\tau_e}(\vec{P}_{0}-\vec{P})
\end{equation}
\noindent Where $\tau_e$ is the electric dipolar relaxation time and $\vec{P}_0$ is the value of polarization would be in equilibrium.\\
\noindent  The usual procedure leads to the following set of equations:\\
\begin{itemize}
\item For the bulk:
\begin{equation}
\frac{\partial\vec{P}^{\pm}}{\partial t}+\vec{\nabla}.(\vec{u}^{\pm}\vec{P}^{\pm}) = \frac{1}{\tau^{\pm}_e}(\vec{P}^{\pm}_{0}-\vec{P}^{\pm})
\end{equation}
\noindent As explained for the magnetization equation, for the bulk phase we can have the following simplified form:
\begin{equation}
\frac{D\vec{P}^{\pm}}{D t} = \frac{1}{\tau^{\pm}_e}(\vec{P}^{\pm}_{0}-\vec{P}^{\pm})
\end{equation}
\item For the interface:
\begin{equation}
\frac{\partial\vec{P}^s}{\partial t}+\vec{\nabla}.(\vec{u}^s\vec{P}^s) + \vec{P}^{+}\vec{n}.(\vec{u}^{+}-\vec{u}^s)-\vec{P}^{-}\vec{n}.(\vec{u}^{-}-\vec{u}^s)-2\mathcal{H}(\vec{u}^s.\vec{n})\vec{P}^s= \frac{1}{\tau^s_e}(\vec{P}^s_{0}-\vec{P}^s)
\end{equation}
\end{itemize}
\noindent As for the magnetization, there is no term leading to supplementary boundary conditions for the polarization. However, let us stress that in usual magnetic fluids the electric polarization is small, and in practice the constitutive equation for the polarization can be reduced to:
\begin{equation}
\vec{P}\approx\vec{P}_0 = \xi_e\varepsilon_0\vec{E}
\end{equation}
\noindent where $\xi_e$ is the electric susceptibility. To obtain the corresponding equations for the bulk phase and for the interface is, in this case, straightforward.
\section{Equations for the energies of the system}
\noindent In the last section we derived the dynamical equations for the physical variables of our system:  the momentum, $\rho\vec{u}$, the electromagnetic field, $\vec{B},\vec{D}$, the angular momentum, $\rho\vec{K}$. Now we want to obtain the corresponding equations for the different kind of energies which exist in our system: we will begin with the kinetic energy and the electromagnetic energy, after that we will consider the equation for the angular momentum energy, and we will finish with the conservation equation for the total and the energy the internal of the system. We have to notice that if we want to do this work we need to reformulate the dynamical equations, using the two following relations (\ref{jj}) and (\ref{jf}), that we reproduce here:
\begin{equation}\label{pre}
\frac{\partial\rho^{\pm}}{\partial t}+\vec{\nabla}.(\rho^{\pm}\vec{u}^{\pm}) = 0
\end{equation}
\begin{equation}\label{seco}
\frac{\partial\rho^{s}}{\partial t}+\vec{\nabla}.(\rho^{s}\vec{u}^{s})+\rho^{+}\vec{n}.(\vec{u}^{+}-\vec{u}^{s})- \rho^{-}\vec{n}.(\vec{u}^{-}-\vec{u}^{s})-2\mathcal{H}(\vec{n}.\vec{u}^s)\rho^s = 0
\end{equation}
With these relations, we will be able to rewrite the equations in a lagrangian form, as we show below. Let's consider the following expression:
\begin{equation}\label{fgh}
\frac{\partial \rho^{s}\vec{F}^{s}}{\partial t}+\vec{\nabla}.(\rho^{s}\vec{F}^{s}\vec{u}^{s})-2\mathcal{H}(\vec{n}.\vec{u}^s)\rho^s\vec{F} = \rho^{s}\left(\frac{\partial \vec{F}^{s}}{\partial t}+\vec{u}^{s}.\vec{\nabla}\vec{F}^{s}\right)+\vec{F}^{s}\left(\frac{\partial \rho^{s}}{\partial t}+\vec{\nabla}.(\rho^{s}\vec{u}^{s})-2\mathcal{H}(\vec{n}.\vec{u}^s)\rho^s)\right)
\end{equation}
This relation hold for any quantity $F$, be it a scalar $F$, a vector $\vec{F}$ or a tensor $\textbf{F}$. Now, if we take into account the relation (\ref{seco}), we obtain from (\ref{fgh}):
 \begin{equation}\label{fgj}
\frac{\partial \rho^{s}\vec{F}^{s}}{\partial t}+\vec{\nabla}.(\rho^{s}\vec{F}^{s}\vec{u}^{s})-2\mathcal{H}(\vec{n}.\vec{u}^s)\rho^s\vec{F} = \rho^{s}\frac{D\vec{F}^s}{Dt}-\vec{F}^{s}\left(\rho^{+}\vec{n}.(\vec{u}^{+}-\vec{u}^{s})- \rho^{-}\vec{n}.(\vec{u}^{-}-\vec{u}^{s})\right)
\end{equation}

\noindent We have a direct relation between the eulerian derivative, $\frac{\partial }{\partial t}$ and the lagrangian derivative ,$\frac{D}{Dt}$, we note that in the last case, the curvature term doesn't appear anymore. We have a similar result for the equation inside the bulk:

\begin{equation}\label{fgp}
\frac{\partial \rho^{s}\vec{F}^{\pm}}{\partial t}+\vec{\nabla}.(\rho^{\pm}\vec{F}^{\pm}\vec{u}^{\pm})= \rho^{\pm}\frac{D\vec{F}^{\pm}}{Dt}
\end{equation}
In the next sections we will  use (\ref{fgj}) and (\ref{fgp}) to transform the equation in a lagrangian form, which will  simplify the calculation for the equations for the different kind of energy.
\subsection{Equations for the kinetic energy}
\noindent With (\ref{fgj}) and (\ref{fgp}), equations (\ref{eru}) and (\ref{ert})  can be rewritten as follow:
\begin{itemize}
\item Equations for the bulks:\\
\begin{equation}
\rho^{\pm}\frac{D\vec{u}^{\pm}}{Dt} = \vec{\nabla}.(\tilde{\textbf{T}}^{\pm}+\textbf{T}^{\pm}_{em}) -\frac{\partial}{\partial t} (\vec{D}\times\vec{B})^{\pm}+\rho^{\pm}\vec{F}_{ex}
 \end{equation}
\item Equations for the interface:\\
\begin{equation*}
\rho^{s}\frac{D_s\vec{u}^{s}}{Dt} = \rho^{-}(\vec{n}.(\vec{u}^{-}-\vec{u}^s))(\vec{u}^{-}-\vec{u}^s) - \rho^{+}(\vec{n}.(\vec{u}^{+}-\vec{u}^s))(\vec{u}^{+}-\vec{u}^s)+\vec{\nabla}.\tilde{\textbf{T}}^{s}+\vec{n}.(\tilde{\textbf{T}}^{+}-\tilde{\textbf{T}}^{-})+\rho^s\vec{F}_{ex}
\end{equation*}
\begin{equation}
\vec{\nabla}.\textbf{T}^{s}_{em}+\vec{n}.(\textbf{T}^{+}_{em}-\textbf{T}^{-}_{em})+\vec{n}.\vec{u}^s\left[(\vec{D}\times\vec{B})^{+}-(\vec{D}\times\vec{B})^{-}\right]-\frac{\partial}{\partial t}(\vec{D}\times\vec{B})^s
\end{equation}
\item boundary conditions:\\
These conditions remain the same as before:
\begin{equation}
\vec{n}.(\tilde{\textbf{T}}^s+\textbf{T}^s_{em})+\vec{u}^s.\vec{n}(\vec{D}\times\vec{B})^s = 0
\end{equation}
\end{itemize}
The equation for the kinetic energy, defined as $\mathcal{U}_{cin} = \frac{1}{2}\vec{u}^2$, can be obtained by taking the scalar product of the last two equations with the vector $\vec{u}$.\\
\noindent
We have for the bulks:
\begin{equation}
\rho^{\pm}\frac{D\vec{u}^{\pm}}{Dt}.\vec{u}^{\pm} = \left(\vec{\nabla}.(\tilde{\textbf{T}}^{\pm}+\textbf{T}^{\pm}_{em}).\vec{u}^{\pm}\right) - \frac{\partial\vec{p}^{\pm}_{em}}{\partial t}.\vec{u}^{\pm}+\rho^{\pm}\vec{F}_{ex}.\vec{u}^{\pm}
\end{equation}
We use use the relation:
\begin{equation}\label{abcd}
(\vec{\nabla}.\textbf{T}).\vec{u} = -\vec{\nabla}.(\textbf{T}\vec{u})+\textbf{T}:\vec{\nabla}\vec{u}
\end{equation}
to obtain the formulation:
\begin{equation}
\rho^{\pm}\frac{D}{Dt}(\frac{(\vec{u}^{\pm})^2}{2}) = -\vec{\nabla}.\left((\tilde{\textbf{T}}^{\pm}+\textbf{T}^{\pm}_{em})\vec{u}^{\pm}\right)+\left(\tilde{\textbf{T}}^{\pm}+\textbf{T}^{\pm}_{em}\right):\vec{\nabla}\vec{u}^{\pm}-\frac{\partial\vec{p}^{\pm}_{em}}{\partial t}.\vec{u}^{\pm}+\rho^{\pm}\vec{F}_{ex}.\vec{u}^{\pm}
\end{equation}

\noindent
At the interface, the equations have the form:
\begin{equation*}
\rho^s\frac{D_s}{Dt}\left(\frac{\vec{u}^s}{2}\right)^2 = \rho^{-}\left[\left(\vec{n}.(\vec{u}^{-}-\vec{u}^s)\right)(\vec{u}^{-}-\vec{u}^{s})\right].\vec{u}^s-\rho^{+}\left[\left(\vec{n}.(\vec{u}^{+}-\vec{u}^s)\right)(\vec{u}^{+}-\vec{u}^s)\right].\vec{u}^s+
\end{equation*}
\begin{equation}
\left(\vec{\nabla}.(\tilde{\textbf{T}}^s+\textbf{T}^s_{em})\right).\vec{u}^s+\left(\vec{n}.(\tilde{\textbf{T}}^{+}-\tilde{\textbf{T}}^{-})\right).\vec{u}^s+\left(\vec{n}.(\textbf{T}^{+}_{em}-\textbf{T}^{-}_{em})\right).\vec{u}^s+(\vec{n}.\vec{u}^s)(\vec{p}^{+}_{em}-\vec{p}^{-}_{em}).\vec{u}^s-\frac{\partial\vec{p}^s_{em}}{\partial t}.\vec{u}^s
\end{equation}
which become, using the same relation (\ref{abcd}):
\begin{equation*}
\rho^s\frac{D_s}{Dt}\left(\frac{\vec{u}^s}{2}\right)^2 = \rho^{-}\left[\left(\vec{n}.(\vec{u}^{-}-\vec{u}^s)\right)(\vec{u}^{-}-\vec{u}^{s})\right].\vec{u}^s-\rho^{+}\left[\left(\vec{n}.(\vec{u}^{+}-\vec{u}^s)\right)(\vec{u}^{+}-\vec{u}^s)\right].\vec{u}^s-
\end{equation*}
\begin{equation*}
\vec{\nabla}.\left((\tilde{\textbf{T}}^s+\textbf{T}^s_{em})\vec{u}^s\right)+\left(\tilde{\textbf{T}}^s+\textbf{T}_{em}\right):\vec{\nabla}\vec{u}^s+\left(\vec{n}.(\tilde{\textbf{T}}^{+}-\tilde{\textbf{T}}^{-})\right).\vec{u}^s+\left(\vec{n}.(\textbf{T}^{+}_{em}-\textbf{T}^{-}_{em})\right).\vec{u}^s+
\end{equation*}
\begin{equation}
(\vec{n}.\vec{u}^s)(\vec{p}^{+}_{em}-\vec{p}^{-}_{em}).\vec{u}^s-\frac{\partial\vec{p}^s_{em}}{\partial t}.\vec{u}^s
\end{equation}
For the next part, it is useful to rewrite the equations in a more conventional form, in term of dissipation functions and fluxes.
\begin{itemize}
\item For the bulk, we have:
\begin{equation}
\frac{\partial \rho^{\pm}\mathcal{U}^{\pm}_{cin}}{\partial t}+\vec{\nabla}.(\rho^{\pm}\vec{u}^{\pm}\mathcal{U}^{\pm}_{cin})+\vec{\nabla}.\vec{J}^{\pm}_{cin}+\sigma^{\pm}_{cin} = 0
\end{equation}
Where we defined the current:
\begin{equation}
\vec{J}^{\pm}_{cin} = (\tilde{\textbf{T}}^{\pm}+\textbf{T}^{\pm}_{em}).\vec{u}^{\pm}
\end{equation}
and the dissipation function:
\begin{equation}
\sigma^{\pm}_{cin} = \frac{\partial\vec{p}^{\pm}_{em} }{\partial t}.\vec{u}^{\pm}-\rho^{\pm}\vec{F}_{ex}.\vec{u}^{\pm}-(\tilde{\textbf{T}}^{\pm}+\textbf{T}^{\pm}_{em}):\vec{\nabla}\vec{u}^{\pm}
\end{equation}
\item For the interface, we obtain:
\begin{equation}
\frac{\partial \rho^{s}\mathcal{U}^{s}_{cin}}{\partial t}+\vec{\nabla}.(\rho^{s}\vec{u}^{s}\mathcal{U}^{s}_{cin})+\vec{\nabla}.\vec{J}^{s}_{cin}+\sigma^{s}_{cin}-2\mathcal{H}(\vec{u}^s.\vec{n})\rho^s\mathcal{U}^{s}_{cin} = 0
\end{equation}
with the current
\begin{equation}
\vec{J}^{s}_{cin} = (\tilde{\textbf{T}}^{s}+\textbf{T}^{s}_{em}).\vec{u}^{s}
\end{equation}
and the dissipation function
\begin{equation*}
\sigma^s_{cin} = \frac{\partial\vec{p}^{s}_{em} }{\partial t}.\vec{u}^{s}-\rho^{s}\vec{F}_{ex}.\vec{u}^{s}-(\tilde{\textbf{T}}^{s}+\textbf{T}^{s}_{em}):\vec{\nabla}\vec{u}^{s}-(\vec{n}.(\tilde{\textbf{T}}^{+}+\textbf{T}^{+}_{em}-(\tilde{\textbf{T}}^{-}+\textbf{T}^{-}_{em}))).\vec{u}^s-(\vec{n}.\vec{u}^s)(\vec{p}^{+}_{em}-\vec{p}^{-}_{em}).\vec{u}^s+
\end{equation*}
\begin{equation}
\mathcal{U}^s_{cin}\vec{n}.(\rho^{-}(\vec{u}^{-}-\vec{u}^s)-\rho^{+}(\vec{u}^{+}-\vec{u}^s))+\vec{u}^s.\left[\rho^{+}(\vec{n}.(\vec{u}^{+}-\vec{u}^s))\vec{u}^{+}-\rho^{-}(\vec{n}.(\vec{u}^{-}-\vec{u}^s))\vec{u}^{-}\right]
\end{equation}
\end{itemize}
The procedure for the next part is the same as here, then we will show directly the results, skipping the details of the calculation.
\subsection{Equations for the electromagnetic energy}
\noindent To obtain the equations for the conservation of the magnetic energy for the bulks and the interface, we will strat from the equations (\ref{ghu}, \ref{huk}) for the bulks and from equations (\ref{dei},\ref{bobu}) for the interface.  
\begin{itemize}
\item Equation for the bulks:\\
\noindent We multiply equation (\ref{ghu}) by $\vec{H}^{\pm}$ and equation (\ref{huk}) by $\vec{E}^{\pm}$, we obtain
\begin{equation}\label{ste1}
\vec{H}^{\pm}.\frac{D\vec{B}^{\pm}}{Dt} = -\vec{H}^{\pm}.(\nabla\times\vec{E}^{\pm})
\end{equation}
\begin{equation}\label{ste2}
\vec{E}^{\pm}.(\nabla\times\vec{H}^{\pm}) = \vec{E}^{\pm}.\frac{D\vec{D}^{\pm}}{Dt}+\vec{j}^{\pm}.\vec{E}^{\pm}
\end{equation}
Substracting equation (\ref{ste2}) from equation (\ref{ste1}) and using the identity $\nabla.(\vec{E}\times\vec{H}) = \vec{H}.(\nabla\times\vec{E})-\vec{E}.(\nabla\times\vec{H})$ we obtain after some manipulations
\begin{equation}
\frac{D(\vec{H}^{\pm}.\vec{B}^{\pm})}{Dt}+\frac{D(\vec{D}^{\pm}.\vec{E}^{\pm})}{Dt}+\nabla.(\vec{E}^{\pm}\times\vec{H}^{\pm})+\vec{j}^{\pm}.\vec{E}^{\pm} = \frac{D\vec{H}^{\pm}}{Dt}.\vec{B}^{\pm}+\vec{D}^{\pm}.\frac{D\vec{E}^{\pm}}{Dt}
\end{equation}
Using the phenomenological relations $\vec{B} = \mu_0(\vec{H}+\vec{M})$ and $\vec{D} = \varepsilon_0\vec{E}+\vec{P}$, we can obtain
\begin{equation}
\frac{D}{Dt}\left[\frac{\mu_0}{2}(H^{\pm})^2+\frac{\varepsilon_0}{2}(E^{\pm})^2\right]+\nabla.(\vec{E}^\pm\times\vec{H}^\pm)+\vec{j}^\pm.\vec{E}^\pm+ \frac{D\vec{P}^{\pm}}{Dt}.\vec{E}^\pm+\mu_0\frac{D\vec{M}^\pm}{Dt}.\vec{H}^\pm = 0
\end{equation}
This equation can be rewritten in an analog form as for the kinetic energy:
\begin{equation}
\frac{\partial \mathcal{U}^{\pm}_{em}}{\partial t}+\nabla.(\vec{u}^{\pm}\mathcal{U}^{\pm}_{em})+\nabla.\vec{J}^{\pm}_{em}+\sigma^{\pm}_{em} = 0
\end{equation}
with the following definition for the electromagnetic energy in the bulk
\begin{equation}
\mathcal{U}^{\pm}_{em} = \frac{\mu_0}{2}(H^{\pm})^2+\frac{\varepsilon_0}{2}(E^{\pm})^2
\end{equation}
and with the dissipation function
\begin{equation}\label{ste3}
\sigma^{\pm}_{em} = \vec{j}^\pm.\vec{E}^\pm+ \frac{D\vec{P}^{\pm}}{Dt}.\vec{E}^\pm+\mu_0\frac{D\vec{M}^\pm}{Dt}.\vec{H}^\pm
\end{equation}
and with the current
\begin{equation}
\vec{J}^{\pm}_{em} = \vec{E}^\pm\times\vec{H}^\pm
\end{equation}
\item Equation for the interface:\\
\noindent We start from the equations (\ref{dei}) and (\ref{bobu}). We follow exactly the same method we used for the bulk phases and we obtain the following equation:
\begin{equation*}
\frac{D_s}{Dt}\left[\frac{\mu_0}{2}(H^{s})^2+\frac{\varepsilon_0}{2}(E^{s})^2\right]-2\mathcal{H}(\vec{u}^s.\vec{n})(\mu_0(H^{s})^2+\varepsilon_0(E^{s})^2)+(\nabla_s.\vec{u}^s)(\mu_0(H^{s})^2+\varepsilon_0(E^{s})^2)+\vec{E}^s.\vec{j}^s+\nabla.(\vec{E}^s\times\vec{H}^s)+
\end{equation*}
\begin{equation*}
 \vec{E}^s.\left(\frac{D_s\vec{P}^s}{Dt}-2\mathcal{H}(\vec{u}^s.\vec{n})\vec{P}^s+\vec{P}^s\nabla_s.\vec{u}^s\right)+ \vec{H}^s.\left(\frac{D_s\vec{M}^s}{Dt}-2\mathcal{H}(\vec{u}^s.\vec{n})\vec{M}^s+\vec{M}^s\nabla_s.\vec{u}^s\right)
\end{equation*}
\begin{equation} 
 +\vec{H}^s.(\vec{n}\times(\vec{E}^{+}-\vec{E}^{-}))-\vec{E}^s.(\vec{n}\times(\vec{H}^{+}-\vec{H}^{-}))+\vec{n}.(\vec{u}^{+}-\vec{u}^s)(\vec{H}^s.\vec{B}^{+}+\vec{E}^s.\vec{D}^{+})-\vec{n}.(\vec{u}^{-}-\vec{u}^s)(\vec{H}^s.\vec{B}^{-}+\vec{E}^s.\vec{D}^{-}) = 0
\end{equation}
Which can be rewritten in the following way
\begin{equation}
\frac{\partial \mathcal{U}^{s}_{em}}{\partial t}+\nabla.(\vec{u}^s\mathcal{U}^s_{em})+\nabla.\vec{J}^{s}_{em}+\sigma^{s}_{em}-2\mathcal{H}(\vec{u}^s.\vec{n})\mathcal{U}^{s}_{em} = 0
\end{equation}
with
\begin{equation}
\mathcal{U}^{s}_{em} = \frac{\mu_0}{2}(H^{s})^2+\frac{\varepsilon_0}{2}(E^{s})^2
\end{equation}
\begin{equation}
\vec{J}^{s}_{em} = \vec{E}^s\times\vec{H}^s
\end{equation}
\begin{equation*}
\sigma^s_{em} = -2\mathcal{H}(\vec{u}^s.\vec{n})\mathcal{U}^{s}_{em}+(\nabla_s.\vec{u}^s)\mathcal{U}^s_{em}+\vec{E}^s.\vec{j}^s+ \vec{E}^s.\left(\frac{D_s\vec{P}^s}{Dt}-2\mathcal{H}(\vec{u}^s.\vec{n})\vec{P}^s+\vec{P}^s\nabla_s.\vec{u}^s\right)+
\end{equation*}
\begin{equation*} \vec{H}^s.\left(\frac{D_s\vec{M}^s}{Dt}-2\mathcal{H}(\vec{u}^s.\vec{n})\vec{M}^s+\vec{M}^s\nabla_s.\vec{u}^s\right)+\vec{H}^s.(\vec{n}\times(\vec{E}^{+}-\vec{E}^{-}))-\vec{E}^s.(\vec{n}\times(\vec{H}^{+}-\vec{H}^{-}))
\end{equation*}
\begin{equation}\label{ste4}
+\vec{n}.(\vec{u}^{+}-\vec{u}^s)(\vec{H}^s.\vec{B}^{+}+\vec{E}^s.\vec{D}^{+})-\vec{n}.(\vec{u}^{-}-\vec{u}^s)(\vec{H}^s.\vec{B}^{-}+\vec{E}^s.\vec{D}^{-}) 
\end{equation}
\item boundary condition: we have the simple following condition
\begin{equation}
\vec{n}.\vec{J}^s_{em}= 0
\end{equation}
\end{itemize}
In the definitions (\ref{ste3}) and (\ref{ste4}) of the dissipation functions appear the total time derivative of the polarization and of the magnetization. In order to obtain a closed set of equations for the electromagnetic energy itself, we can use the constitutive equations for the polarization and for the magnetization.    
\subsection{Equations for the rotational kinetic energy}
\noindent To obtain this equation we follow the same procedure as we did to obtain the equation for the kinetic energy.
We define the internal energy for the rotation as ($\mathcal{U}_{rot} = \frac{1}{2}\vec{\Omega}^2$). The calculation leads to the following results:
\noindent For the bulk
\begin{equation}
\rho^{\pm}\frac{1}{2}\frac{D(\Omega^{\pm})^2}{Dt} = (\vec{\nabla}.\textbf{Y}^{\pm}).\vec{\Omega}^{\pm}-\left(\vec{\textbf{Z}}: (\tilde{\textbf{T}}^{\pm}+\textbf{T}^{\pm}_{em})\right).\vec{\Omega}^{\pm}+ \rho^{\pm}\mathcal{L}.\vec{\Omega}^{\pm}
\end{equation}
\noindent For the interface, we find the following expression:
\begin{equation*}
\rho^{s}\frac{1}{2}\frac{D_s(\Omega^{s})^2}{Dt} = (\vec{\nabla}.\textbf{Y}^{s}).\vec{\Omega}^{s}-\left(\vec{\textbf{Z}}: (\tilde{\textbf{T}}^{s}+\textbf{T}^{s}_{em})\right).\vec{\Omega}^{s}+ \rho^{s}\mathcal{L}.\vec{\Omega}^{s}+\left(\vec{n}.(\textbf{Y}^{+}-\textbf{Y}^{-})\right).\vec{\Omega}^s+
\end{equation*}
\begin{equation}
 \left((\vec{\Omega}^{-}-\vec{\Omega}^{s}).\vec{\Omega}^s\right)\left(\rho^{-}\vec{n}.(\vec{u}^{-}-\vec{u}^s)\right)- \left((\vec{\Omega}^{+}-\vec{\Omega}^{s}).\vec{\Omega}^s\right)\left(\rho^{+}\vec{n}.(\vec{u}^{+}-\vec{u}^s)\right)
\end{equation}
In term of dissipation function and flux, these equations become
\begin{itemize}
\item For the bulk:
\begin{equation}
\frac{\partial\rho^{\pm}\mathcal{U}^{\pm}_{rot}}{\partial t}+\vec{\nabla}.(\rho^{\pm}\mathcal{U}^{\pm}_{rot}\vec{u}^{\pm})+\vec{\nabla}.\vec{J}^{\pm}_{rot}+\sigma^{\pm}_{rot} = 0
\end{equation}
with the current:
 \begin{equation}
 \vec{J}^{\pm}_{rot} = \textbf{Y}^{\pm}.\vec{\Omega}^{\pm}
\end{equation}
with the dissipation function
\begin{equation}
\sigma^{\pm}_{rot} = \left(\vec{\textbf{Z}}:(\tilde{\textbf{T}}^{\pm}+\textbf{T}^{\pm}_{em})\right).\vec{\Omega}^{\pm}-\textbf{Y}^{\pm}:\vec{\nabla}\vec{\Omega}^{\pm}-\rho^{\pm}\mathcal{L}.\vec{\Omega}^{\pm}
\end{equation}
\item For the interface, we have:
\begin{equation}
\frac{\partial\rho^{s}\mathcal{U}^{s}_{rot}}{\partial t}+\vec{\nabla}.(\rho^{s}\mathcal{U}^{s}_{rot}\vec{u}^{s})+\vec{\nabla}.\vec{J}^{s}_{rot}+\sigma^{s}_{rot}-2\mathcal{H}(\vec{n}.\vec{u}^s)\rho^s\mathcal{U}^{s}_{rot} = 0
\end{equation}
with the current
 \begin{equation}
 \vec{J}^{s}_{rot} =\textbf{Y}^{s}.\vec{\Omega}^{s}
\end{equation}
and the dissipation function has the following form
\begin{equation*}
\sigma^s_{rot} = \left(\vec{\textbf{Z}}:(\tilde{\textbf{T}}^{s}+\textbf{T}^{s}_{em})\right).\vec{\Omega}^{s}-\textbf{Y}^{s}:\vec{\nabla}\vec{\Omega}^{s}-\rho^{s}\mathcal{L}.\vec{\Omega}^{s}-\left(\vec{n}.(\textbf{Y}^{+}-\textbf{Y}^{-})\right).\vec{\Omega}^s-
\end{equation*}
\begin{equation}
\vec{\Omega}^s.\left[\vec{\Omega}^{-}\rho^{-}\vec{n}.(\vec{u}^{-}-\vec{u}^s)-\vec{\Omega}^{+}\rho^{+}\vec{n}.(\vec{u}^{+}-\vec{u}^s)\right]+\mathcal{U}^s_{rot}\vec{n}.(\vec{u}^{-}-\vec{u}^s)-\mathcal{U}^s_{rot}\rho^{+}\vec{n}.(\vec{u}^{+}-\vec{u}^s)
\end{equation}
\end{itemize}
\section{Conservation of the total energy and internal energy}
\noindent We start with the total energy, $\rho\mathcal{E}$, which is defined as
\begin{equation}
\rho\mathcal{E} = \rho\mathcal{U}_{in}+\rho\mathcal{U}_{cin}+\rho\mathcal{U}_{rot}+\mathcal{U}_{em}
\end{equation}
This quantity is conserved, therefore the conservation equation has the form:
\begin{equation}\label{last}
\frac{\partial \rho\mathcal{E}}{\partial t}+\vec{\nabla}.(\rho\mathcal{E}\vec{u})+\vec{\nabla}.\vec{J}_e = 0
\end{equation}
with the current flux, $\vec{J}_e$:
\begin{equation}
\vec{J}_e = \vec{J}_q +\vec{J}_{cin}+\vec{J}_{em}+\vec{J}_{rot}
\end{equation}
Now, we can rewrite it as:
\begin{equation}
\frac{D}{Dt}\int{\rho\mathcal{E}}{dV_T} = \int{\vec{\nabla}.\vec{J}_e}{dV_T}
\end{equation}

\noindent Applying a similar method as we did for the equation for the  mass conservation  we have the following results: for the bulks we have the well known equations:
\begin{equation}
\frac{\partial \rho^{\pm}\mathcal{E}^{\pm}}{\partial t} +\vec{\nabla}.(\rho^{\pm}\vec{u}^{\pm}\mathcal{E}^{\pm})+\vec{\nabla}.\vec{J}^{\pm}_e = 0
\end{equation}
This equation is just the equation (\ref{last}), with the bulk's variables. For the interface, the equation takes the following form:
\begin{equation}
\frac{\partial\rho^s \mathcal{E}^s}{\partial t} +\vec{\nabla}.\vec{J}^s_e+\mathcal{E}^{+}\vec{n}.(\vec{u}^{+}-\vec{u}^s)-\mathcal{E}^{-}\vec{n}.(\vec{u}^{-}-\vec{u}^s)+\vec{n}.(\vec{J}^{+}_e-\vec{J}^{-}_e)-2\mathcal{H}(\vec{n}.\vec{u}^s)\rho^s\mathcal{E}^s+\vec{\nabla}.(\rho^s\vec{u}^s\mathcal{E}^s)= 0
\end{equation}
with the boundary condition:
\begin{equation}
\vec{J}_e.\vec{n} = 0
\end{equation}
\noindent To obtain the equation for the internal energy, $\rho\mathcal{U}_{in}$, we need to start from the equation for the total energy and substract from it the expressions of the equations for the other quantities, which leads to the following expression:
\begin{equation}
\frac{\partial \rho\mathcal{U}_{in}}{\partial t}+\vec{\nabla}.(\rho\vec{u}\mathcal{U}_{cin})+\vec{\nabla}.\vec{J}_q = \sigma_{em}+ \sigma_{cin}+ \sigma_{rot}
\end{equation}
The use of the usual decomposition procedure leads to the following results:
\begin{itemize}
\item For the bulks:
\begin{equation}
\frac{\partial \rho^{\pm}\mathcal{U}^{\pm}_{in}}{\partial t}+\vec{\nabla}.(\rho^{\pm}\vec{u}^{\pm}\mathcal{U}^{\pm}_{cin})+\vec{\nabla}.\vec{J}^{\pm}_q = \sigma^{\pm}_{em}+ \sigma^{\pm}_{cin}+ \sigma^{\pm}_{rot}
\end{equation}
\item For the interface:
\begin{equation}
\frac{\partial \rho^{s}\mathcal{U}^{s}_{in}}{\partial t}+\vec{\nabla}.(\rho^{s}\vec{u}^{s}\mathcal{U}^{s}_{cin})+\vec{\nabla}.\vec{J}^{s}_q+\vec{n}.\vec{u}^s(\rho^{-}\mathcal{U}^{-}_{in}-\rho^{+}\mathcal{U}^{+}_{in})+\vec{n}.(\rho^{+}\vec{u}^{+}\mathcal{U}^{+}_{in}-\rho^{-}\vec{u}^{-}\mathcal{U}^{-}_{in}) = \sigma^{s}_{em}+ \sigma^{s}_{cin}+ \sigma^{s}_{rot}+2\mathcal{H}(\vec{n}.\vec{u}^s)\rho^s\mathcal{U}^{s}_{in}
\end{equation}
\item Boundary condition for the current
\begin{equation}
\vec{n}.\vec{J}^s_q = 0
\end{equation}
\end{itemize}
\section{Entropy production and fluxes}
\noindent In this  last section we consider the entropy production of our system. The first step will be the derivation of the equation for the evolution of the entropy and in a second step we will compute the different fluxes. As we did before we can write the equation for the entropy, $\mathcal{S}$ in a integral form:
\begin{equation}
\frac{D}{Dt}\int{\rho\mathcal{S}}{dV}+\int{\vec{\nabla}.\vec{J}_s}{dV} = \int{\Phi_s}{dV}
\end{equation}
Applying the usual method, we obtain the following set of equations:
\begin{itemize}
\item For the bulk:
\begin{equation}\label{abcd}
\frac{\partial \rho^{\pm}\mathcal{S}^{\pm}}{\partial t}+\vec{\nabla}.(\vec{J}^{\pm}_s+\rho^{\pm}\vec{u}^{\pm}\mathcal{S}^{\pm}) = \Phi^{\pm}_s
\end{equation}
\item For the interface:
\begin{equation}\label{efg}
\frac{\partial \rho^{s}\mathcal{S}^{s}}{\partial t}  +\vec{\nabla}.(\vec{J}^{s}_s+\rho^{s}\vec{u}^{s}\mathcal{S}^{s})-2\mathcal{H}(\vec{u}^s.\vec{n})\rho^s\mathcal{S}^s=\Phi^{s}_s+ \rho^{-}\mathcal{S}^{-}\vec{n}.(\vec{u}^{-}-\vec{u}^s)-\rho^{+}\mathcal{S}^{+}\vec{n}.(\vec{u}^{+}-\vec{u}^s)
\end{equation}
\item boundary condition:
we have
\begin{equation}
\vec{n}.\vec{J}^s_s = 0
\end{equation}
\end{itemize}
We have to specify the entropy currents $\vec{J}^{\pm}_s$ and $\vec{J}^s_s$ and also the dissipation functions $\Phi^s_s$ and $\Phi^{\pm}_s$. We start with the well known relation between the internal energy and the entropy:\\
\noindent For the bulk:
\begin{equation}
\rho^{\pm}\frac{D\mathcal{S}^{\pm}}{Dt} = \frac{\rho^{\pm}}{T^{\pm}}\frac{D\mathcal{U}^{\pm}_{in}}{Dt}-\frac{p^{\pm}}{\rho^{\pm}T^{\pm}}\frac{D\rho^{\pm}}{Dt}+\frac{\vec{E}^{\pm}}{T^{\pm}}.\frac{D\vec{P}^{\pm}}{Dt}+\mu_0\frac{\vec{H}^{\pm}}{T^{\pm}}.\frac{D\vec{M}^{\pm}}{Dt}
\end{equation}
\noindent For the interface:
\begin{equation}
\rho^{s}\frac{D_s\mathcal{S}^s}{Dt} = \frac{\rho^{s}}{T^{s}}\frac{D_s\mathcal{U}^s_{in}}{Dt}-\frac{p^s}{\rho^{s}T^{s}}\frac{D_s\rho^{s}}{Dt}+\frac{\vec{E}^{s}}{T^{s}}.\frac{D_s\vec{P}^{s}}{Dt}+\mu_0\frac{\vec{H}^{s}}{T^{s}}.\frac{D_s\vec{M}^{s}}{Dt}
\end{equation}
Using the definition of the lagrangian derivative, we have:\\
\noindent For the bulk:
\begin{equation}
\frac{\partial\rho^{\pm}\mathcal{S}^{\pm}}{\partial t}+ \vec{\nabla}.(\rho^{\pm}\vec{u}^{\pm}\mathcal{S}^{\pm}) = \frac{1}{T^{\pm}}\left[\frac{\partial\rho^{\pm} \mathcal{U}^{\pm}_{in}}{\partial t}+\vec{\nabla}.(\rho^{\pm}\vec{u}^{\pm}\mathcal{U}^{\pm}_{in})\right]-\frac{p^{\pm}}{\rho^{\pm}T^{\pm}}\frac{D\rho^{\pm}}{Dt}+\frac{\vec{E}^{\pm}}{T^{\pm}}.\frac{D\vec{P}^{\pm}}{Dt}+\mu_0\frac{\vec{H}^{\pm}}{T^{\pm}}.\frac{D\vec{M}^{\pm}}{Dt}
\end{equation}
\noindent For the interface:
\begin{equation*}
\frac{\partial\rho^{s}\mathcal{S}^{s}}{\partial t}+ \vec{\nabla}.(\rho^{s}\vec{u}^{s}\mathcal{S}^{s})-\mathcal{S}^s\rho^{-}\vec{n}.(\vec{u}^{-}-\vec{u}^s)+\mathcal{S}^s\rho^{+}\vec{n}.(\vec{u}^{+}-\vec{u}^s) =
\end{equation*}
\begin{equation*}
\frac{1}{T^{s}}\left[\frac{\partial\rho^{s} \mathcal{U}^{s}_{in}}{\partial t}+\vec{\nabla}.(\rho^{s}\vec{u}^{s}\mathcal{U}^{s}_{in})-2\mathcal{H}(\vec{n}.\vec{u}^s)\rho^s\mathcal{U}^s_{in}-\mathcal{U}^s_{in}\rho^{-}\vec{n}.(\vec{u}^{-}-\vec{u}^s)+\mathcal{U}^s_{in}\rho^{+}\vec{n}.(\vec{u}^{+}-\vec{u}^s)\right]
\end{equation*}
\begin{equation}
-\frac{p^{s}}{\rho^{s}T^{s}}\frac{D_s\rho^{s}}{Dt}+\frac{\vec{E}^{s}}{T^{s}}.\frac{D_s\vec{P}^{s}}{Dt}+\mu_0\frac{\vec{H}^{s}}{T^{s}}.\frac{D_s\vec{M}^{s}}{Dt}
\end{equation}
Using the equations for the internal energy we obtain\\
\noindent For the bulk:
\begin{equation}
\frac{\partial\rho^{\pm}\mathcal{S}^{\pm}}{\partial t}+ \vec{\nabla}.(\rho^{\pm}\vec{u}^{\pm}\mathcal{S}^{\pm}) = \frac{1}{T^{\pm}}\left[-\vec{\nabla}.\vec{J}^{\pm}_q + \sigma^{\pm}_{em}+ \sigma^{\pm}_{cin}+ \sigma^{\pm}_{rot}\right]-\frac{p^{\pm}}{\rho^{\pm}T^{\pm}}\frac{D\rho^{\pm}}{Dt}+\frac{\vec{E}^{\pm}}{T^{\pm}}.\frac{D\vec{P}^{\pm}}{Dt}+
\end{equation}
\begin{equation*}
\mu_0\frac{\vec{H}^{\pm}}{T^{\pm}}.\frac{D\vec{M}^{\pm}}{Dt}
\end{equation*}
\noindent For the interface:
\begin{equation*}
\frac{\partial\rho^{s}\mathcal{S}^{s}}{\partial t}+ \vec{\nabla}.(\rho^{s}\vec{u}^{s}\mathcal{S}^{s})-\mathcal{S}^s\rho^{-}\vec{n}.(\vec{u}^{-}-\vec{u}^s)+\mathcal{S}^s\rho^{+}\vec{n}.(\vec{u}^{+}-\vec{u}^s) =
\end{equation*}
\begin{equation*} \frac{1}{T^{s}}\left[-\vec{\nabla}.\vec{J}^{s}_q-\vec{n}.\vec{u}^s(\rho^{-}\mathcal{U}^{-}_{in}-\rho^{+}\mathcal{U}^{+}_{in})-\vec{n}.(\rho^{+}\vec{u}^{+}\mathcal{U}^{+}_{in}-\rho^{-}\vec{u}^{-}\mathcal{U}^{-}_{in})\right] +
\end{equation*}
\begin{equation*}
 \frac{1}{T^{s}}\left[\sigma^{s}_{em}+ \sigma^{s}_{cin}+ \sigma^{s}_{rot}-2\mathcal{H}(\vec{n}.\vec{u}^s)\rho^s\mathcal{U}^s_{in}-\mathcal{U}^s_{in}\rho^{-}\vec{n}.(\vec{u}^{-}-\vec{u}^s)+\mathcal{U}^s_{in}\rho^{+}\vec{n}.(\vec{u}^{+}-\vec{u}^s)\right]
\end{equation*}
\begin{equation}
-\frac{p^{s}}{\rho^{s}T^{s}}\frac{D_s\rho^{s}}{Dt}+\frac{\vec{E}^{s}}{T^{s}}.\frac{D_s\vec{P}^{s}}{Dt}+\mu_0\frac{\vec{H}^{s}}{T^{s}}.\frac{D_s\vec{M}^{s}}{Dt}
\end{equation}
Now, we can take into account the following relation:
\begin{equation}
\frac{1}{T} \vec{\nabla}.\vec{J}_q = \vec{\nabla}.(\frac{\vec{J}_q}{T})+\frac{1}{T^2}\vec{J}_q.\vec{\nabla} T
\end{equation}
then we can obtain the equation:\\
\noindent For the bulk:
\begin{equation*}
\frac{\partial\rho^{\pm}\mathcal{S}^{\pm}}{\partial t}+ \vec{\nabla}.(\rho^{\pm}\vec{u}^{\pm}\mathcal{S}^{\pm})+\vec{\nabla}.(\frac{\vec{J}^{\pm}_q}{T^{\pm}}) = \frac{1}{T^{\pm}}\left[-\frac{1}{T^{\pm}}(\vec{\nabla} T).\vec{J}^{\pm}_q + \sigma^{\pm}_{em}+ \sigma^{\pm}_{cin}+ \sigma^{\pm}_{rot}\right]-\frac{p^{\pm}}{\rho^{\pm}T^{\pm}}\frac{D\rho^{\pm}}{Dt}
\end{equation*}
\begin{equation}\label{der}
+\frac{\vec{E}^{\pm}}{T^{\pm}}.\frac{D\vec{P}^{\pm}}{Dt}+\mu_0\frac{\vec{H}^{\pm}}{T^{\pm}}.\frac{D\vec{M}^{\pm}}{Dt}
\end{equation}
comparison with (\ref{abcd}) leads to the following identification:
\begin{equation}
\vec{J}^{\pm}_s = \frac{\vec{J}^{\pm}_q}{T^{\pm}}
\end{equation}
The left side of (\ref{der}) can be seen as the dissipation function $\Phi^{\pm}_S$.\\
\noindent For the interface:
\begin{equation*}
\frac{\partial\rho^{s}\mathcal{S}^{s}}{\partial t}+ \vec{\nabla}.(\rho^{s}\vec{u}^{s}\mathcal{S}^{s})-\mathcal{S}^s\rho^{-}\vec{n}.(\vec{u}^{-}-\vec{u}^s)+\mathcal{S}^s\rho^{+}\vec{n}.(\vec{u}^{+}-\vec{u}^s)+\vec{\nabla}.(\frac{\vec{J}^s_q}{T^s}) =
\end{equation*}
\begin{equation*}
\frac{1}{T^{s}}\left[-\frac{1}{T^s}(\vec{\nabla} T).\vec{J}^{s}_q-\vec{n}.\vec{u}^s(\rho^{-}\mathcal{U}^{-}_{in}-\rho^{+}\mathcal{U}^{+}_{in})-\vec{n}.(\rho^{+}\vec{u}^{+}\mathcal{U}^{+}_{in}-\rho^{-}\vec{u}^{-}\mathcal{U}^{-}_{in})\right]
\end{equation*}
\begin{equation*}
 + \frac{1}{T^{s}}\left[-2\mathcal{H}(\vec{u}^s\vec{n})\rho^s\mathcal{U}^s_{in}+\sigma^{s}_{em}+ \sigma^{s}_{cin}+ \sigma^{s}_{rot}-\mathcal{U}^s_{in}\rho^{-}\vec{n}.(\vec{u}^{-}-\vec{u}^s)+\mathcal{U}^s_{in}\rho^{+}\vec{n}.(\vec{u}^{+}-\vec{u}^s)\right]
\end{equation*}
\begin{equation}
-\frac{p^{s}}{\rho^{s}T^{s}}\frac{D_s\rho^{s}}{Dt}+\frac{\vec{E}^{s}}{T^{s}}.\frac{D_s\vec{P}^{s}}{Dt}+\mu_0\frac{\vec{H}^{s}}{T^{s}}.\frac{D_s\vec{M}^{s}}{Dt}
\end{equation}
we can compare this expression with (\ref{efg}), and we obtain
\begin{equation}
\vec{J}^{s}_s = \frac{\vec{J}^{s}_q}{T^{s}}
\end{equation}

\noindent The left side is the function $\Phi^{s}_S+2\mathcal{H}\rho^s\mathcal{S}^s(\vec{u}^s.\vec{n})$. In the expressions we have obtained here, some terms which are function of the mean curvature appear, there are absent from the previous work for the same reason as explained before. If we compare our results with the results in \cite{re}, we notice some differences coming from  the fact that in \cite{re}, the drift terms, as we said before, have been neglected. Thus,the expressions we have obtained here are a generalization of the previous results.\\

\section{Conclusion}
\noindent In this paper, we have shown how the effect of the time dependent curvature can be taken into account in the framework of the theory built by D. Bedeaux, A.M. Albano and their coworkers. We obtained very general equations for a system of two immiscible liquid phases with an interface, for which system, the presence of singular electromagnetic field is taken into account. In the limit of the simplifications which were done in previous works, we recover the results of \cite{ra}$-$\cite{re}. We have also stressed  some features of the equations in their lagrangian form for which the non linear term disappears leading to simpler but complete equations. We end up with a unified theory for the mathematical treatment of such systems, since various previous approaches have been shown to lead to identical results. In future works, we plan to develop the Onsager relations and to show what is exactly the effect of this non linear term. Indeed, most authors have not given  any illustration of the effect of this term or explained how it should be or not important.
\section*{Acknowledgments}\noindent
Q. Vanhaelen thanks F.R.I.A (Fond pour la formation ˆ la recherche dans l'industrie et dans l'agriculture) for its financial support.  Support provided in the frame of the Agreement for Scientific Cooperation between the Bulgarian Academy of Sciences, Bulgaria and the WBI (Wallonie-Brussels International, former name: Commissariat General des Relations Internationales, French Community of Belgium) is gratefully acknowledged.
\pagebreak

\end{document}